\documentstyle[12pt]{article}

\catcode`\@=11

   \renewcommand{\section}%
   {\setcounter{equation}{0}\@startsection {section}{1}{\z@}{-3.5ex plus -1ex
   minus -.2ex}{2.3ex plus .2ex}{\Large\bf}}

\catcode`\@=12

\def\C{{\rm\kern.24em \vrule width.02em height1.4ex depth-.05ex \kern-.26em
C}}
\def\R{{\rm I\kern-.20em R}}
\def\F{{\rm I\kern-.20em F}}
\def\P{{\rm I\kern-.20em P}}
\def\N{{\rm I\kern-.20em N}}
\def\Q{{\rm\kern.24em \vrule width.02em height1.4ex depth-.05ex \kern-.26em
Q}}
\def\pa{\partial}

\newcommand{\beq}{\begin{equation}}
\newcommand{\eeq}{\end{equation}}

\begin{document}
\baselineskip=20pt
\pagestyle{plain}

\title{Topics in Quantum Geometry \\ of Riemann Surfaces:\\
Two-Dimensional Quantum Gravity}
\author{
Leon A.~Takhtajan \\
Department of Mathematics\\
SUNY at Stony Brook\\
Stony Brook, NY 11794-3651 \\
U.S.A.}
\date{}
\maketitle

\section{Beginning Concepts}

\subsection{Introduction}

In these lectures, we present geometric approach to the two-dimensional
quantum gravity. It became popular since Polyakov's discovery that
first-quantized bosonic string propagating in $\R^d$ can be described as
theory of $d$ free bosons coupled with the two-dimensional quantum
gravity~\cite{Pol1}. In critical dimension $d=26$, the gravity
decouples and Polyakov's approach reproduces results obtained earlier
by different methods (see, e.g., \cite{AG1} for detailed discussion and
references).

Classically, the two-dimensional gravity is a theory formulated on a smooth
oriented two-dimensional surface $X$, endowned with a Riemannian metric $ds^2$,
whose dynamical variables are metrics in the conformal class of $ds^2$.
Classical equation of motion is the two-dimensional Einstein equation with a
cosmological term and it describes the metric with constant Gaussian curvature.
Since in two dimensions, conformal structure uniquely determines a complex
structure, a surface $X$ with the conformal class of $ds^2$ has a structure of
a one-dimensional complex manifold---a Riemann surface. We will consider the
case when $X$ is either compact (i.e.~an algebraic curve), or it is
non-compact, having finitely many branch points of infinite order. Except for
few cases, the
virtual Euler characteristic $\chi(X)$ of the Riemann surface $X$ is negative
so that, according to the Gauss-Bonnet theorem, it admits metrics with constant
negative curvature only. Specifically, if $\chi(X)<0$, then there exists on $X$
a unique complete conformal metric of constant negative curvature $-1$, called
the Poincar\'{e}, or hyperbolic, metric. In terms of local complex coordinate
$z$ on $X$, conformal metric has the form $ds^2=e^{\phi}|dz|^2$ and its
Gaussian curvature is given by $R_{ds^2}=-2e^{-\phi}\phi_{z \bar{z}}$, where
subscripts indicate partial derivatives. The condition of constant negative
curvature $-1$ is equivalent to the following nonlinear PDE,
\beq \label{LE}
\frac{\pa^2 \phi}{\pa z \pa \bar{z}}=\frac{1}{2}e^{\phi},
\eeq
called Liouville equation.

Quantization of the two-dimensional gravity (in the conformal gauge) amounts
to the quantization of conformal metrics with the classical action
given by the Liouville theory. The definition of the latter is a non-trivial
problem. Namely, since $\phi(z, \bar{z})$ is not a globally defined function
on $X$, but rather a logarithm of the conformal factor of the metric,
``kinetic term'' $|\phi_z|^2 dz \wedge d\bar{z}$ does not yield a
$(1,1)$-form on $X$ and, therefore, can not be integrated over $X$. This means
that ``naive'' Dirichlet' type functional is not well defined and can not
serve as an action for the Liouville theory (it also diverges at the branch
points). There are two possible ways to deal with this situation. The first one
uses the choice of the background metric on $X$, whereas in the second, a
regularization at the branch points is used. In addition, when topological
genus of $X$ is zero, one also takes advantage of the existence of a single
global coordinate on $X$~\cite{ZT1}; for the case of the non-zero genus one
uses a global coordinate, provided by the Schottky uniformization~\cite{ZT2}.
In the first approach the rich interplay between semi-classical approximation,
conformal symmetry and the uniformization of Riemann surfaces seems to be lost,
or at least hidden. In the second approach, developed in~\cite{ZT1,ZT2} (see
also~\cite{T3} for a review), and which we will use here, this interplay plays
a fundamental role.

Once the action functional is defined, one can quantize the two-dimensional
gravity using a method of functional integration, performing the
``summation'' over all conformal metrics on the Riemann surface $X$, with a
hyperbolic metric being a ``critical point'' of the ``integral''. One may
refer to it as to the ``quantization'' of the hyperbolic geometry of Riemann
surfaces, as fluctuations around the Poincar\'{e} metric ``probe'' the
hyperbolic geometry. In other words, the two-dimensional quantum gravity
can be considered as a special topic of the ``quantum geometry'' of Riemann
surfaces. Among other things, it provides unified treatment of conformal
symmetry, uniformization and complex geometry of moduli spaces.

The following argument illustrates our approach. Let $X$ be two-dimensional
sphere $S^2$, realized as Riemann sphere $\P^1$ with complex coordinate
$z$ in standard chart $\C$. A smooth conformal metric on $X$ has the form
$ds^2=e^{\phi}|dz|^2$ with condition $\phi(z, \bar{z}) \simeq -4 \log|z|$
as $z \rightarrow \infty$, ensuring the regularity at $\infty$. Regularized
action functional is defined as
\beq \label{NLA}
S_0(\phi)=\lim_{R \rightarrow \infty}(\int_{|z| \leq R}
(|\phi_z|^2 +e^{\phi})d^{2}z-8\pi \log R),~~d^{2}z=\frac{\sqrt{-1}}{2}dz \wedge
d\bar{z},
\eeq
and corresponding Euler-Lagrange equation $\delta S_0=0$ yields the Liouville
equation. However, it has no global smooth solution, since sphere $S^2$,
according to Gauss-Bonnet, admits only metrics of constant positive
curvature. At the quantum level one should consider functional integral
$$<S^2>={\cal \int} {\cal D}\phi~e^{-(1/2 \pi h) S_0(\phi)},$$
where ``integration'' goes over all smooth conformal metrics on $S^2$,
${\cal D} \phi$ symbolizes certain ``integration measure''
$${\cal D}\phi \doteq \prod_{x \in S^2}d\phi(x),$$
and positive $h$ plays the role of a coupling constant. According to the
latter remark, this functional integral does not have critical points, so that
the perturbation theory, based on the ``saddle point method'', is not
applicable. However, quantity $<S^2>$ has a meaning of a
partition function/ground state energy of the theory, and as such plays a
normalization role only. Objects of fundamental importance are given by the
correlation functions of Liouville vertex operators $V_{\alpha}(\phi)(z)=
\exp\{\alpha \phi(z, \bar{z})\}$ with different
``charges'' $\alpha$. Namely, according to Polyakov~\cite{Pol1}, these
correlation functions, which should be calculated in order to find the
scattering amplitudes of non-critical strings, are given by the following
functional integral
\beq \label{NFI}
<V_{\alpha_1}(z_1) \cdots V_{\alpha_n}(z_n)>={\cal \int} {\cal D}\phi~
V_{\alpha_1}(\phi)(z_1) \cdots V_{\alpha_n}(\phi)(z_n)e^{-(1/2 \pi h)
S_0(\phi)}.
\eeq
Introducing sources $\delta(z-z_i)$, localized at insertion
points $z_i$, one can include the product of the vertex operators into the
exponential of the action, so that it acquires a critical point, given by a
singular metric on $S^2$. The main proposal of Polyakov~\cite{Pol2} is that
the ``summation'' over smooth metrics with the insertion of vertex
operators in (\ref{NFI}) should be equivalent to the ``summation'' over
metrics with singularities at the insertion points, without the insertion of
the vertex operators! For special $\alpha$'s these singular metrics become
complete metrics on Riemann surface $X$---Riemann sphere $\P^1$ with branch
points $z_i$. If $\chi(X)$ is negative, a complete hyperbolic metric on $X$
exists and perturbation theory is applicable (see also~\cite{Sei} for similar
arguments).

Specifically, consider branch points $z_i$ of orders $2 \leq l_i \leq \infty$,
$i=1, \ldots, n$. According to Poincar\'{e}~\cite{P1} (see
also~\cite[pp.~72-78]{Kra}), admissible singularities
of metric $ds^2=e^{\phi(z,\bar{z})}|dz|^2$ are of the following form
\beq \label{ES}
e^{\phi} \simeq l^{-2}_{i}\frac{r_{i}^{2/l_i-2}}{(1-r_{i}^{1/l_i})^2},
\eeq
when $l_i < \infty$, and
\beq \label{PS}
e^{\phi} \simeq \frac{1}{r_{i}^{2} \log^{2}r_i},
\eeq
when $l_i=\infty$ and $r_i \doteq |z-z_i| \rightarrow 0$. Note that the metric
$$\frac{|dz|^2}{|z|^2 \log^{2}|z|},$$
used in (\ref{PS}), is the Poincar\'{e} metric on the punctured unit disc
$\{z \in \C~|~0<|z|<1\}$.
Assuming for a moment that $\infty$ is a regular point, one also has
$\phi \simeq -4 \log|z|$, as $z \rightarrow \infty$.
Using the formula
$$\frac{\pa^2}{\pa z \pa \bar{z}} \log|z|^2=\pi\delta(z),$$
where $\delta(z)$ is the Dirac delta-function, equation (\ref{LE}) with
asymptotics (\ref{ES})---(\ref{PS}) can be rewritten in the following form
\beq \label{LES}
\frac{\pa^2 \phi}{\pa z \pa \bar{z}}=\frac{1}{2}e^{\phi}-\pi \sum_{i=1}^{n}
(1-1/l_i)\delta(z-z_i) + 2 \pi \delta(1/z),
\eeq
where insertion points are present explicitly (cf.~\cite{Sei}).
Equation (\ref{LES}) is uniquely solvable if
\beq \label{GB}
\chi(X)=2-\sum_{i=1}^{n}(1-1/l_i) <0,
\eeq
i.e.~if one can ``localize curvature'' at branch points. As we shall
see later, corresponding charges are
\beq \label{VC}
\alpha_i=\frac{1}{2h}(1-1/l_{i}^{2}),
\eeq
and play a special role in the Liouville theory. Our main interest will be
concentrated on the case when branch points $z_i$ are of infinite order,
i.e.~$l_i=\infty$. In this case Riemann surface $X$ is non-compact, being
a Riemann sphere $\P^1$ with $n$ removed distinct points, called punctures.
According to (\ref{PS}), punctures are ``points at infinity'' in the
intrinsic geometry on $X$, defined by metric $e^{\phi}|dz|^2$, i.e.~for any
$z_0 \in X$ the geodesic distance $d(z_0,z) \rightarrow \infty$ as $z
\rightarrow z_i$.

This approach does not look like a standard one in the quantum field theory;
however, the Liouville theory is not a standard model either! It provides
a manifestly geometrical treatment of the theory, as one should expect from
the theory of gravity. On the other hand, the standard approach to the
two-dimensional quantum gravity~\cite{KPZ,D,DK} (see also~\cite{DH} for a
review), which is based on the free-field representation, does not recover
the underlying hyperbolic geometry at a classical limit. These two approaches,
perhaps, might reveal the two phases of the two-dimensional quantum gravity;
relation between them is yet to be discovered.

Consistent perturbative treatment of the geometrical approach, given
in~\cite{T1,T2}, exhibits the richness of the theory. Mathematically, it
provides a unified view on the uniformization of Riemann surfaces and complex
geometry of moduli spaces, and uses methods of the Teichm\"{u}ller theory,
spectral geometry and theory of automorphic forms. Physically, it realizes
conformal bootstrap program of Belavin-Polyakov-Zamolodchikov~\cite{BPZ}
in the geometrical setting of Friedan-Shenker~\cite{FS}. In particular,
conformal Ward identities can be proved, and the central charge of the
Virasoro algebra can be calculated. Detailed exposition of these results
will be presented in the course, which will be organized as follows.

In \S1.2 we formally define the main objects of the theory: the expectation
value of Riemann surfaces and normalized connected forms of multi-point
correlation functions with the stress-energy tensor components. In
\S\S2.1---2.3 we recall basic facts from the conformal field theory, adapted
to our case: operator product expansion of Belavin-Polyakov-Zamolodchikov,
conformal Ward identities and Friedan-Shenker geometric interpretation. In \S
\S3.1---3.2 we give a perturbative definition of the expectation value and
correlation functions; in \S3.3 we compute one - and two-point correlation
functions at the tree level and in the one-loop approximation. In \S4.1 we
show how conformal Ward identities yield new mathematical results on K\"{a}hler
geometry of moduli spaces; in \S4.2 we discuss corresponding results for
the one-loop approximation and compute the central charge of the Virasoro
algebra. In \S4.3 we recall basic facts from the Teichm\"{u}ller theory,
and in \S4.4 we show how to prove the results, obtained in \S \S4.1---4.2.
In particular, validity of the conformal Ward identities for the
quantum Liouville theory follows. Finally, in \S4.5 we briefly discuss how to
adapt our
approach to the case of Riemann surfaces of non-zero genus. All sections
are accompanied by exercises, intended to develop the familiarity
with various aspects of the mathematical formalism used in this course.

\subsection{Quantization of the Hyperbolic Geometry}

Let $X$ be an $n$-punctured sphere, i.e.~Riemann sphere $\P^{1}$ with $n$
removed distinct points $z_1, \ldots, z_n$, called punctures. Without loss of
generality we assume $z_n=\infty$, so that $X=\C \setminus \{z_1, \ldots,
z_{n-1} \}$. Using single global complex coordinate $z$ on $\C$, any
conformal metric on $X$ can be represented as
$$ds^2=e^{\phi(z, \bar{z})}|dz|^{2}.$$
Denote by ${\cal C}(X)$ a class of all smooth conformal metrics on $X$ having
asymptotics (\ref{PS}), where for $i=n$ ($z_n=\infty$ is now a puncture!) one
has $r_n=|z| \doteq r \rightarrow \infty$. These asymptotics ensure that
${\cal C}(X)$ consists of complete metrics on $X$, which are
asymptotically hyperbolic at the punctures. Asymptotics (\ref{PS}) imply that
the standard expression (\ref{NLA}) for the Liouville action diverges when
$\phi \in {\cal C}(X)$. Properly regularized Liouville action was presented
in~\cite{ZT1} and has the form
\beq \label{LA}
S(\phi)=
\lim_{\epsilon \rightarrow 0} \{  \int_{X_{\epsilon}}(|\phi_{z}|^{2}
+e^{\phi})d^{2}z +2\pi n \log\epsilon +4\pi (n-2)\log|\log\epsilon| \},
\eeq
where $X_{\epsilon}=
X \setminus  \bigcup^{n-1}_{i=1}\{r_i<\epsilon\}
\bigcup \{r>1/\epsilon\}$. Euler-Lagrange equation $\delta S=0$ yields
Liouville equation ---the equation for complete conformal metric on $X$ of
constant negative curvature $-1$. When virtual Euler characteristic $\chi(X)$
of Riemann surface $X$ is negative, i.e.~when, according to (\ref{GB}),
$n \geq 3$, Liouville equation  has a unique solution, denoted by $\phi_{cl}$.

Following the discussion in the previous section, the correlation functions of
the puncture operators (correlation functions of the punctures), ``symbolized''
in (\ref{NFI}), should be depicted by the following functional integral
\beq \label{FI}
<X>={\cal \int}_{{\cal C}(X)} {\cal D}\phi~e^{-(1/2 \pi h)S(\phi)}.
\eeq
We call $<X>$ the ``expectation value'' of Riemann surface $X$ and
consider it as the main object of the two-dimensional quantum gravity, which
encodes all the information about the theory. Being unable to give a
nonperturbative definition, we define $<X>$ using the perturbation expansion
around classical solution $\phi= \phi_{cl}$.
(Similar approach was proposed in~\cite{Zam} for the case when $\phi_{cl}$
corresponds to the Fubini-Study metric on $\P^1$). We describe this procedure
in \S3.1.

Classical Liouville theory is conformally invariant: this is another way of
saying that dynamical variables are conformal metrics $ds^2$ on Riemann
surface $X$. This invariance implies, in particular, that the stress-energy
tensor of the Liouville theory, which ``measures'' the respond of the theory
to the local deformations of the metric, is traceless. Its $(2,0)$-component
$T(\phi)(z)$ is given by
\beq \label{SET}
T(\phi)=\frac{1}{h}(\phi_{zz}-\frac{1}{2}\phi_{z}^{2}),
\eeq
is conserved on classical equations of motion
\beq \label{CL}
\partial_{\bar{z}} T_{cl}=0,
\eeq
where $T_{cl}(z)=T(\phi_{cl})(z)$, and has the transformation law of a
projective connection (times $1/h$) under holomorphic change of coordinates,
i.e.
\beq \label{PC}
\tilde{T}(w)=T(f(w))f^{\prime}(w)^{2}+
\frac{1}{h}{\cal S}(f)(w),
\eeq
where $z=f(w)$. Here ${\cal S}$ stands for the Schwarzian derivative of
(locally) holomorphic function $f$,
$${\cal S}(f)=\frac{f^{\prime \prime \prime}}{f^{\prime}}-
\frac{3}{2} (\frac{f^{\prime \prime}}{f^{\prime}})^{2},$$
which satisfies transformation law
\beq \label{CI}
{\cal S}(f \circ g)={\cal S}(f) \circ g(g^{\prime})^2 + {\cal S}(g),
\eeq
called the Cayley identity.

{}From (\ref{PS}) it follows that $T(\phi)$ has a second order poles at
punctures
\beq \label{SOP1}
T(\phi)(z)=\frac{1}{2h(z-z_i)^2} + o(|z-z_i|^{-2}),~{\rm as}~z \rightarrow
z_i,~i=1, \ldots, n-1,
\eeq
and
\beq \label{SOP2}
T(\phi)(z)=\frac{1}{2hz^2} + o(|z|^{-2}),~{\rm as}~z \rightarrow \infty.
\eeq
Since $T_{cl}$ is holomorphic on $X$, we also have
\beq \label{CSET}
T_{cl}(z)=\sum_{i=1}^{n-1}(\frac{1}{2h(z-z_i)^2}+\frac{c_i}{z-z_i}),
\eeq
where, in virtue of (\ref{SOP2}),
\beq \label{C}
\sum_{i=1}^{n-1}c_i=0,~~\sum_{i=1}^{n-1}z_i c_i=1-\frac{n}{2}.
\eeq
As we shall see in \S4.1, expansion (\ref{CSET}) is ultimately related
with the Fuchsian uniformization of Riemann surface $X$.

The $(0,2)$-component $\bar{T}(\phi)(z)$ of the stress-energy tensor
is given by
$$\bar{T}(\phi)=\frac{1}{h}(\phi_{\bar{z}\bar{z}}-\frac{1}{2}\phi_{\bar{z}}^{2}
)$$
and has similar properties.

We emphasize that the ``modification'' of the stress-energy tensor,
i.e.~addition of total derivative $\phi_{zz}/h$ to ``free-field'' term
$-\phi_{z}^{2}/2h$ in (\ref{SET}), is a crucial feature of the theory. It is
an artifact of the property that $\phi$  is not a scalar field, but transforms
like a logarithm of the conformal factor of metric under the holomorphic
change of coordinates
\beq \label{ConF}
\tilde{\phi}(w,\bar{w})=\phi(f(w),\overline{f(w)}) + \log|f^{\prime}(w)|^2,~
z=f(w).
\eeq
Therefore the naive ``free-field'' expression for the stress-energy tensor has
a wrong transformation law and only the addition of a total derivative term
yields the correct transformation law (\ref{PC}). This modification of the
stress-energy tensor goes back to Poincar\'{e}~\cite{P1}.

Multi-point correlation functions of the holomorphic and anti-holomorphic
components of the stress-energy tensor in the presence of punctures
are defined by the following functional integral
\beq \label{CF}
<\prod_{i=1}^{k}T(z_i) \prod_{j=1}^{l}\bar{T}(\bar{w}_j)X> \doteq
{\cal \int}_{{\cal C}(X)} {\cal D}\phi~\prod_{i=1}^{k}T(\phi)(z_i)\prod_{j=1}^{
l}\bar{T}(\phi)(\bar{w}_j)~ e^{-(1/2\pi h) S(\phi)}.
\eeq
We present detailed prescription for calculating these correlation functions
in Section 3.2. Here, following the standards of quantum field theory, we
introduce a normalized connected (reduced) form of these correlation functions.
Denote by $I$ the set $\{z_1, \ldots, z_k, \bar{w}_1, \ldots, \bar{w}_l \}$,
so that correlation function (\ref{CF}) can be simply depicted as $<I>$. The
normalized connected form of (\ref{CF}), which we denote by
\beq \label{NCF}
<<\prod_{i=1}^{k}T(z_i) \prod_{j=1}^{l}\bar{T}(\bar{w}_j)X>> \doteq <<I>>,
\eeq
is defined by the following inductive formula
\beq \label{DNCF}
<<I>>=\frac{<I>}{<X>}~-~\sum_{r=2}^{k+l} \sum_{I=I_1 \cup \cdots I_r}
<<I_1>> \cdots <<I_r>>,
\eeq
where summation goes over all representations of set $I$ as a disjoint union
of subsets $I_1, \ldots, I_r$. Thus, for example,
\begin{eqnarray*}
<<T(z)X>> &=& \frac{<T(z)X>}{<X>},~~~~~~~~<<\bar{T}(\bar{z})X>>~=~
\frac{<\bar{T}(\bar{z})X>}{<X>}, \\
<<T(z)T(w)X>> &=& \frac{<T(z)T(w)X>}{<X>}~-~<<T(z)X>><<T(w)X>>,\\
<<T(z)\bar{T}(\bar{w})X>> &=& \frac{<T(z)\bar{T}(\bar{w})X>}{<X>}~-~<<T(z)X>>
<<\bar{T}(\bar{w})X>>.
\end{eqnarray*}

Next, following Schwinger~\cite{JS} (see e.g.~also~\cite[Ch.~7.3]{Huan}) we
present a generating functional for the correlation functions with
stress-energy tensor components. In our geometrical setting, Schwinger's
external sources are represented by Beltrami differentials on Riemann surface
$X$. Recall that Beltrami differential $\mu$ is a tensor of the type $(-1,1)$
on $X$, i.e.~it has a transformation
law
\beq \label{BD}
\tilde{\mu}(w,\bar{w})=\mu(f(w),\overline{f(w)})\frac{\overline{f^{\prime}(w)}}
{f^{\prime}(w)},
\eeq
under the holomorphic change of coordinates $z=f(w)$. In addition, Beltrami
differentials are bounded in the $L^{\infty}$ norm:
$$||\mu||_{\infty}\doteq \sup_{z \in X}|\mu(z,\bar{z})| < \infty.$$
The generating functional for the normalized multi-point correlation functions
of the stress-energy tensor is introduced by the following expression
\beq \label{z}
{\cal Z}(\mu,\bar{\mu};X)=\frac{Z(\mu,\bar{\mu};X)}{<X>},
\eeq
where
\beq \label{Z}
Z(\mu,\bar{\mu};X) \doteq {\cal \int}_{{\cal C}(X)}{\cal D}\phi \exp\{-\frac{1}
{2 \pi h}S(\phi) + {\rm v.p.}\int_{X}(T(\phi)\mu + \bar{T}(\phi)\bar{\mu})
d^{2}z\}.
\eeq
Here, since $X$ admits a single global coordinate, the $(2,0)$-component of
the stress-energy tensor ($(0,2)$-component) can be considered as quadratic
differential $Tdz^2$ ($\bar{T}d\bar{z}^2$), so that $T\mu dz \wedge d \bar{z}$
($\bar{T}\bar{\mu} dz \wedge d\bar{z}$) is a $(1,1)$-form on $X$ and can be
integrated over $X$. The integral in (\ref{Z}) is understood in the principal
value sense in virtue of singularities (\ref{SOP1})---(\ref{SOP2}).

Generating functional for the normalized connected multi-point correlation
functions is defined in a standard fashion
\beq \label{W}
\frac{1}{h} {\cal W}(\mu,\bar{\mu};X)=\log {\cal Z}(\mu, \bar{\mu};X),
\eeq
so that
\beq \label{CCF}
h<<\prod_{i=1}^{k}T(z_i) \prod_{j=1}^{l}\bar{T}(\bar{w}_j)X>>=
\frac{\delta^{k+l}{\cal W}(\mu,\bar{\mu};X)}{\delta \mu(z_1) \cdots \delta \mu
(z_k)\delta \bar{\mu}(\bar{w}_1) \cdots \delta \bar{\mu}(\bar{w}_l)}|_{\mu=\bar
{\mu}=0}.
\eeq

In \S3.2  we present a formalism for calculating the generating functional
${\cal W}$ in all orders of the perturbation theory.

{\bf Problems}

{\bf 1} Evaluate explicitly the integral of the Gaussian curvature of the
conformal metric on $S^2$ and show that $S^2$ does not admit a smooth constant
negative curvature metric.

{\bf 2} Prove all properties of the stress-energy tensor, mentioned in the
lecture: the transformation law (\ref{PC}), conservation on classical solution
(\ref{CL}), and representation (\ref{CSET}).

{\bf 3} Consider as an example, formal power series $A$ and $B$ in
infinitely many variables $x_i, i \in \N$,
$$A=\sum_{I}a_I x_I,~B=\log A=\sum_I b_I x_I,$$
where $I=\{i_1, \ldots, i_l\}$ runs through all finite subsets of the
set $\N$ of natural numbers, and $x_I=x_{i_1} \cdots x_{i_l}$, and prove that
normalized connected multi-point correlation functions, defined by formula
(\ref{CCF}), satisfy relation (\ref{DNCF}).

\section{Conformal Symmetry}
\subsection{Operator Product Expansion}

The basic fact that $ds^2=e^{\phi}|dz|^2$ is a conformal metric on
Riemann surface $X$, can be reformulated by saying that positive
quantity $e^{\phi(z,\bar{z})}$ has transformation properties of a
$(1,1)$-tensor on $X$. This implies that at the classical level, vertex
operators $V_{\alpha}(\phi)(z,\bar{z})$ are tensors of type
$(\alpha, \alpha)$ on $X$, i.e.~they satisfy transformation law
$$\tilde{V}_{\alpha}(\tilde{\phi})(w,\bar{w})=V_{\alpha}(\phi)(f(w),
\overline{f(w)})|f^{\prime}(w)|^{2\alpha},$$
under holomorphic change of coordinates $z=f(w)$. Therefore, they can be
considered as primary fields with classical conformal dimensions (weights)
$\Delta =\bar{\Delta}=\alpha$.

One should expect that at the quantum level conformal symmetry of the theory
is preserved, so that vertex operators and components of the stress-energy
tensor satisfy operator product expansions (OPE) of
Belavin-Polyakov-Zamolodchikov~\cite{BPZ}
\begin{eqnarray} \label{OPE}
T(z)T(w)&=&\frac{c/2}{(z-w)^4}+\frac{2T(w)}{(z-w)^2}+\frac{1}{z-w}~\frac{\pa T}
{\pa w}(w)+ {\rm regular~terms}, \\
\bar{T}(\bar{z})\bar{T}(\bar{w})&=&\frac{c/2}{(\bar{z} - \bar{w})^4}+
\frac{2\bar{T}(\bar{w})}{(\bar{z}-\bar{w})^2} + \frac{1}{\bar{z}-\bar{w}}~
\frac{ \pa \bar{T}}{\pa \bar{w}}(\bar{w}) + {\rm regular~terms}, \\
T(z)\bar{T}(\bar{w})&=&{\rm regular~terms},
\end{eqnarray}
and
\begin{eqnarray} \label{TV}
T(z)V_{\alpha}(w, \bar{w})&=&\frac{\Delta_{\alpha}}{(z-w)^2}V_{\alpha}(w,
\bar{w}) + \frac{1}{z-w}~\frac{\pa V_{\alpha}}{\pa w}(w, \bar{w}) +{\rm
regular~terms},\\
\bar{T}(\bar{z})V_{\alpha}(w, \bar{w})&=&\frac{\bar{\Delta}_{\alpha}}
{(\bar{z}-\bar{w})^2} V_{\alpha}(w, \bar{w})+\frac{1}{\bar{z}-\bar{w}}~
\frac{\pa V_{\alpha}}{\pa\bar{w}}(w, \bar{w}) + {\rm regular~terms},
\end{eqnarray}
where $c=c(h)$ is the central charge of the theory, $\Delta_{\alpha}=
\bar{\Delta}_{\alpha}= \Delta_{\alpha}(h)$ are conformal dimensions of
Liouville vertex operators, and ``regular terms'' denote expressions, which
are regular as $z \rightarrow w$. These OPE are understood as exact expansions
of the correlation functions~\cite{BPZ}.

Using perturbation theory, we will prove the validity of (2.1)---(2.5) and
calculate the value of $c$, which turns out to be $1+12/h$. We will also show
that conformal dimensions of geometric vertex operators coincide with
their classical charges. Actually, the latter result can be verified directly,
using the identification between correlation function (\ref{NFI}) of
vertex operators and expectation value (\ref{FI}). Namely, consider the
thrice-punctured sphere with punctures at $z_1, z_2$ and $z_3=\infty$.
It is isomorphic (global conformal symmetry) to the normalized sphere with
punctures $0,1,\infty$. This isomorphism is given by the
fractional-linear transformation $w=z_1-z/z_1-z_2$; using (\ref{ConF})
it is easy to see that
$$e^{\tilde{\phi}(w)}=|z_1-z_2|^2e^{\phi(z)},$$
and (cf.~\cite{Zog})
$$S(\phi)=S(\tilde{\phi})+ 2\pi \log|z_1-z_2|.$$
Performing the change of variables $\phi \mapsto \tilde{\phi}$ in functional
integral (\ref{FI}), we get
\beq \label{CD}
<X>_{z_1,z_2,\infty}=\frac{<X>_{0,1,\infty}}{|z_1-z_2|^{1/h}},
\eeq
which shows that scaling dimension $d=\Delta + \bar{\Delta}$ of the
puncture operator is $1/h$, so that $\Delta=1/2h$. Indeed, $<X>_{z_1,z_2,
\infty}$ should be considered as a three-point correlation function of
the puncture operators $V_{\alpha}(\phi)(z)$, inserted at points
$z_1,z_2$ and $\infty$. According to~\cite{BPZ}, the three-point correlation
function has the form
$$<V_{\alpha}(z_1)V_{\alpha}(z_2)V_{\alpha}(z_3)>=\frac{C}{|z_1-z_2|^d
|z_1-z_3|^d |z_2-z_3|^d}.$$
Since $V_{\alpha}(\infty)=\lim_{z \rightarrow \infty}|z|^{-2d}V_{\alpha}(1/z)$,
we get from (\ref{CD}) that $d=1/h$. (Note that if one interpetes $<X>_{z_1,
z_2,\infty}$ as a two-point correlation function, one gets a wrong value
$1/2h$ for the scaling dimension of the puncture operator!).

It is also instructive to compare the Liouville theory with other solvable
models of the conformal field theory, like minimal models of BPZ
and WZW model. It is well-known that the latter are examples of the
so-called rational conformal field theories, which have the property that
``physical'' correlation functions of primary fields are finite sums of the
products of the holomorphic and anti-holomorphic conformal blocks. Quantum
groups naturally enter into this picture, with $R$-matrices describing
the monodromy of the conformal blocks. In the case of the quantum Liouville
theory, one can also expect the chiral splitting of expectation
value $<X>$ into a ``sum'' (or rather an integral)
$$<X>=\sum_{\beta}{\cal F}_{\beta}(X)\overline{{\cal F}_{\beta}(X)}$$
of the products of the holomorphic and anti-holomorphic conformal blocks.
However, such decomposition is not yet known. Possible role of quantum groups
in the quantum geometry, as well as the role of Poisson-Lie groups in the
hyperbolic geometry, is yet another problem to be investigated.

\subsection{Conformal Ward Identities}

According to BPZ~\cite{BPZ}, conformal symmetry of the  theory, expressed
through OPE, can be equivalently formulated in terms of the infinite sequence
of conformal Ward identities (CWI), that relate correlation functions
involving stress-energy tensor components with correlation functions
without them. Specifically, OPE (2.1) in the presence of punctures can be
rewritten as
\beq \label{WIT}
<T(z)X>=\sum_{i=1}^{n-1}(\frac{\Delta}{(z-z_i)^2}+\frac{1}{z-z_i}~\frac{\pa}
{\pa z_i}) <X>,
\eeq
where $\Delta$ stands for the conformal dimension of the puncture operator.
Similarly, OPE (2.1) yields the following CWI
\begin{eqnarray} \label{WITT}
<T(z)T(w)X>&=&\frac{c/2}{(z-w)^4}<X>+\{\frac{2}{(z-w)^2}+\frac{1}{z-w}~\frac
{\pa} {\pa w} \nonumber \\
& + & \sum_{i=1}^{n-1}(\frac{\Delta}{(z-z_i)^2}+\frac{1}{z-z_i}~\frac{\pa}
{\pa z_i})\}<T(w)X>.
\end{eqnarray}
Correlation functions, involving $(0,2)$-component $\bar{T}$ of the
stress-energy tensor, satisfy similar CWI (see, e.g., \cite{Dot}).

Global ${\rm PSL}(2,\C)$-symmetry of the theory also imposes constraints
on correlation functions~\cite{BPZ}. Fixing $z_n=\infty$ reduces this symmetry
to the invariance under translations and dilations, and yields the following
equations
\beq \label{GWI}
\sum_{i=1}^{n-1}\frac{\pa}{\pa z_i}<X>=0,~~\sum_{i=1}^{n-1}(z_i \frac{\pa}
{\pa z_i}+\Delta)<X>= \Delta<X>,
\eeq
(cf.~relations (\ref{C})), so that
\beq \label{TAI}
<T(z)X>=\frac{\Delta}{z^2}<X> + O(|z|^{-3}),~~z \rightarrow \infty
\eeq
(cf.~asymptotics (\ref{SOP2})).

Using (\ref{GWI}), one can fix the global conformal symmetry by setting
$z_{n-2}=0, z_{n-1}=1, z_n=\infty$ and obtain normalized form of CWI.
Namely, denoting by ${\cal L} (z)$ and $\bar{{\cal L}}(\bar{z})$ the following
first order differential operators
\beq \label{L}
{\cal L}(z)=\sum_{i=1}^{n-3}R(z,z_i)\frac{\pa}{\pa z_i},~
\bar{{\cal L}}(\bar{z})=\sum_{i=1}^{n-3}R(\bar{z},\bar{z}_i)
\frac{\pa}{\pa \bar{z}_i},
\eeq
where
\beq \label{R}
R(z,z_i)=\frac{1}{z-z_i}+\frac{z_i-1}{z}-\frac{z_i}{z-1}=\frac{z_i(z_i-1)}
{z(z-1)(z-z_i)},
\eeq
we can rewrite CWI (\ref{WIT}), and its complex conjugate, in the following
succinct form
\begin{eqnarray}\label{NT}
<<T(z)X>>_0 & \doteq & <<T(z)X>>-~T_{s}(z)={\cal L}(z)\log<X>, \\
<<\bar{T}(\bar{z})X>>_0 & \doteq & <<\bar{T}(\bar{z})X>>-~\bar{T}_{s}(\bar{z})=
\bar{{\cal L}}(\bar{z}) \log<X>,
\end{eqnarray}
where
\beq \label{TS}
T_{s}(z)=\sum_{i=1}^{n-1}\frac{\Delta}{(z-z_i)^2}+\frac{(2-n)\Delta}
{z(z-1)}~.
\eeq
CWI (\ref{WITT}), and its complex conjugate, can be written as follows
\begin{eqnarray}  \label{NTT}
<<T(z)T(w)>> & = & \frac{c/2}{(z-w)^4} \nonumber \\
&+&\{ 2R_{w}(z,w)+R(z,w) \frac{\pa}{\pa w}+ {\cal L}(z) \} <<T(w)X>>, \\
<<\bar{T}(\bar{z})\bar{T}(\bar{w})>> & = & \frac{c/2}{(\bar{z}-\bar{w})^4}
 \nonumber \\
&+&\{ 2R_{\bar{w}}(\bar{z},\bar{w})+R(\bar{z},\bar{w})\frac{\pa}{\pa \bar{w}} +
\bar{{\cal L}}(\bar{z})\}<<\bar{T}(\bar{w})X>>.
\end{eqnarray}
Finally,
\beq \label{NTCT}
<<T(z)\bar{T}(\bar{w})X>>={\cal L}(z)<<\bar{T}(\bar{w})X>>=
{\cal L}(z)\bar{{\cal L}}(\bar{w})\log<X>.
\eeq

According to BPZ~\cite{BPZ}, relations (\ref{NT})---(\ref{NTCT}) state, at the
level of correlation functions, that puncture operators are primary fields and
$T(z), \bar{T}(\bar{z})$ are generating functions of the holomorphic and
anti-holomorphic Virasoro algebras, that mutually commute in virtue of
(\ref{NTCT}).

Similar formulas can be obtained for multi-point correlation functions.
Namely, introducing the following first-order differential operators
$$D(z,w)=R(z,w)\frac{\pa}{\pa w}  + 2R_w(z,w),~\bar{D}(\bar{z}, \bar{w})=
R(\bar{z},\bar{w})\frac{\pa}{\pa \bar{w}}+2R_{\bar{w}}(\bar{z}, \bar{w}),$$
and using notation (cf. \S1.2)
$$<<1,\ldots, k;\overline{1, \ldots,l}>> \doteq
<<\prod_{i=1}^{k}T(z_i)\prod_{j=1}^{l}\bar{T}(\bar{w}_j)X>>,$$
we get for $k+l>2$
\beq \label{MT}
<<1, \ldots, k; \overline{1, \ldots, l}>>=
(\sum_{i=2}^{k}D(z_1,z_i)+{\cal L}(z_1))<<2, \ldots, k; \overline{1, \ldots,
l}>>,
\eeq
\beq \label{MCT}
<<1, \ldots, k; \overline{1, \ldots, l}>>=
(\sum_{j=2}^{l}\bar{D}(\bar{w}_1,\bar{w}_j)+\bar{\cal L}(\bar{w}_1))
<<1, \ldots, k; \overline{2, \ldots, l}>>.
\eeq
These formulas, combined with (\ref{NT})---(\ref{NTCT}), express multi-point
correlations functions with stress-energy tensor components through expectation
value $<X>$.

According to BPZ, constraints imposed by CWI (together with constraints
from possible additional symmetries) allow to solve the theory completely. In
most interesting examples (minimal models of BPZ, WZW model) this is indeed
the case and complete solution can be obtained by representation theory (of
the Virasoro algebra, Kac-Moody algebra). The simplest case of the minimal
models of BPZ corresponds to the discrete series representations of the
Virasoro algebra (see~\cite{BPZ,F} and~\cite{G} for a review).

In our geometrical approach, correlation functions are defined through a
functional integral, and we need to affirm the validity of
(\ref{NT})---(\ref {NTCT}). This will be done in \S4, thus providing dynamical
proof of the conformal symmetry. Note that in our formulation, we do not use
representation theory of the Virasoro algebra and calculate the central charge
and conformal dimensions through CWI. In doing so, we tacitly assume that
certain analog of the ``reconstruction theorem'' exists, so that one may indeed
talk about Virasoro algebra representations involved. It is interesting to
understand their realization, as well as the structure of corresponding
conformal blocks. Contrary to the case of minimal models, where algebraic
constructions based on the Verma modules have been used, in case of the
two-dimensional gravity constructions should be geometrical. Since in our case
$c>1$, we should have ``principal series'' representations, as opposed to the
discrete series for minimal models with $c<1$.

Conformal Ward identities (\ref{NT})---(\ref{NTCT}) (more precisely, OPE
(2.1)---(2.5)) can be written as a single universal Ward identity for
generating functional ${\cal W}$, and formally coincides with the Ward
identity in the light-cone gauge, derived by Polyakov~\cite{Pol3} (see
also~\cite{FS,V}). Namely, we have identities
\begin{eqnarray} \label{UW1}
(\frac{\pa}{\pa \bar{z}}+\pi \mu \frac{\pa}{\pa z}+2 \pi \mu_z)\frac{\delta
{\cal W}} {\delta \mu(z)}(\mu,\bar{\mu};X) \nonumber \\
=-\frac{\pi hc}{12} \mu_{zzz} + \frac{\pa}{\pa \bar{z}}(hT_{s}(z)+{\cal L}(z))
\{{\cal W}(\mu, \bar{\mu};X)+h \log<X>\},
\end{eqnarray}
and
\begin{eqnarray} \label{UW2}
(\frac{\pa}{\pa z} + \pi \bar{\mu}\frac{\pa}{\pa \bar{z}} +2 \pi \bar{\mu}_{
\bar{z}}) \frac{\delta{\cal W}}{\delta \bar{\mu}(z)}(\mu,\bar{\mu};X)
\nonumber \\
=-\frac{\pi hc}{12} \bar{\mu}_{\bar{z} \bar{z} \bar{z}} + \frac{\pa}{\pa z}
(h\bar{T}_{s}(\bar{z})+\bar{{\cal L}}(\bar{z}))\{{\cal W}(\mu,\bar{\mu};X)+
h\log<X>\},
\end{eqnarray}
which should be understood at the level of generating functions.
Using formula
$$\frac{\pa}{\pa \bar{z}}R(z,w)=\pi \delta(z-w),$$
where $z,w \neq 0,1$, one easily gets (\ref{NT})---(\ref{NTCT}) from
(\ref{UW1})---(\ref{UW2}).

\subsection{Friedan-Shenker Modular Geometry}

Denote by ${\cal M}_{0,n}$ the moduli space of Riemann surfaces of genus $0$
with $n > 3$ punctures. It can be obtained as a quotient of the space of
punctures $Z_n= \{(z_1, \ldots, z_{n-3}) \in \C^{n-3}~|~z_i \neq 0,1~
{\rm and}~ z_i \neq z_j~{\rm for}~i \neq j \}$ by the action of a symmetric
group of $n$ elements
$${\cal M}_{0,n} \simeq Z_n/S_n.$$
Here symmetric group $S_n$ acts on $Z_n$ as a permutation of the $n$-tuple
$(z_1, \ldots, z_{n-3},0,1,\infty)$, followed by the component-wise action of
the ${\rm PSL}(2,\C)$, normalizing (if necessary) the last three components
back to $0,1, \infty$. As will be explained in \S4.3, moduli space
${\cal M}_{0,n}$ is a complex orbifold of complex dimension $n-3$ and
admits a natural K\"{a}hler structure.  Denote by  $d$
the exterior differential on the spaces ${\cal M}_{0,n}$ and $Z_n$; it has a
standard  decomposition $d=\pa+\bar{\pa}$, where
$$\pa=\sum_{i=1}^{n-3}\frac{\pa}{\pa z_i} dz_i,~\bar{\pa}=\sum_{i=1}^{n-3}
\frac{\pa}{\pa \bar{z}_i}d\bar{z}_i.$$
As we shall see in \S4.3, vectors $\pa/\pa z_i$, which form a basis of
the (holomorphic) tangent space to $Z_n$ at point $(z_1, \ldots, z_{n-3})$,
corresponding to Riemann surface $X$, can be represented by harmonic
Beltrami differentials on $X$. Corresponding (holomorphic) cotangent space at
point $X$ can be identified with a linear space of harmonic quadratic
differentials on $X$. Remarkably, the dual basis to $\pa/\pa z_i$, which
consists of $(1,0)$-forms $dz_i$, admits an explicit description on Riemann
surface $X$. Namely, introducing
\beq \label{PQD}
P_i(z)=-\frac{1}{\pi}R(z,z_i),~i=1,\ldots, n-3,
\eeq
we can identify $(1,0)$-forms $dz_i$ on $Z_n$ with quadratic differentials
$P_i(z)dz^2$ on $X$, so that under this identification (see~\cite{ZT1} and
\S4.3)
\beq \label{D}
\pa=-\frac{1}{\pi}{\cal L}(z)dz^2,~\bar{\pa}=-\frac{1}{\pi}\bar{\cal L}
(\bar{z})d\bar{z}^2.
\eeq

In~\cite{FS}, Friedan-Shenker envisioned a  general ``philosophy'' of the
modular geometry, which describes conformal theories in two dimensions in
terms of a complex geometry of projective bundles (possibly
infinite-dimensional) over moduli spaces of Riemann surfaces. In particular,
according to the ideology in~\cite{FS}, expectation value $<X>$ should
be interpreted as a Hermitian metric in a certain (holomorphic projective)
line bundle over ${\cal M}_{0,n}$, and quadratic differential
$<<T(z)X>>_{0}dz^2$---as a $(1,0)$-component of a canonical metric connection.
Using correspondence (\ref{D}), we can read (\ref{NT}) as
\beq \label{NTF}
<<T(z)X>>_{0}dz^2=-\frac{1}{\pi}\partial \log<X>,
\eeq
in perfect agreement with~\cite[formula (6)]{FS}!

Similarly, the Ward identity (\ref{NTCT}) can be rewritten as
\beq \label{KM}
<<T(z)\bar{T}(\bar{w})X>>dz^2d\bar{w}^{2}=\frac{1}{\pi^2}\partial
\bar{\partial} \log<X>,
\eeq
which allows to interpret $<<T(z)\bar{T}(\bar{w})X>>$ as a curvature form of
the canonical metric connection (cf.~\cite[formula (15)]{FS}). As we
shall see in \S\S4.1---4.2, these formulas encode remarkable relations
between quantum Liouville theory and the K\"{a}hler geometry of moduli
space ${\cal M}_{0,n}$, and provide modular geometry of Friedan-Shenker with
a meaningful example.

However, so far our arguments were rather formal, since we did not define
rigorously our main objects: the expectation value and correlation functions.
This will be done in \S\S3.1---3.2.

{\bf Problems}

{\bf 1} (Research problem) Describe the chiral splitting of the correlation
functions of puncture operators, conformal blocks, their monodromy, etc. What
role do the quantum groups play in this approach?

{\bf 2} Prove CWI (\ref{NT})---(\ref{NTCT}) for normalized connected
correlation functions and (\ref{MT})---(\ref{MCT}) for multi-point correlation
functions.

{\bf 3} Derive universal Wards identities (\ref{UW1})---(\ref{UW2}) from
OPE (2.1)---(2.5).

{\bf 4} Show that differential operator $\pa^{3}_{z} + \mu \pa_z +2 \mu_z$,
where $\mu$ is a Beltrami differential and which appears in
(\ref{UW1})---(\ref{UW2}), maps projective connections into $(2,1)$-tensors on
$X$.

\section{Expectation Value and Correlation Functions}

\subsection{Expectation Value $<X>$}

We define expectation value $<X>$ using the perturbation expansion of
functional integral (\ref{FI}) around classical solution $\phi=\phi_{cl}$.
This expansion will be understood in the sense of formal Laurent series in
$h$, thus defining $\log<X>$ as following
$$\log<X> \doteq N-\frac{1}{2 \pi h}S_{cl}+\sum_{l=0}^{\infty}{\cal X}_l h^l.$$
Here $N$ is an overall infinite constant, that does not depend on $z_i$ and
drops out from all normalized correlation functions, and $S_{cl}=S(\phi_{cl})$
is a classical Liouville action, i.e.~the critical value of the action
functional. Coefficients ${\cal X}_l$---``higher loop contributions''---are
given by the following procedure.

Start with the expansion of the Liouville action around classical solution
\beq \label{LAE}
S(\phi_{cl} + \delta \phi)-S(\phi_{cl})=\int_{X}\delta \phi(L_0 + 1/2)
(\delta \phi)d\rho + \frac{1}{6}\int_{X} (\delta \phi)^{3}d\rho + \cdots, \eeq
where $d\rho=e^{\phi_{cl}}d^2z$ is a volume form of the Poincar\'{e} metric,
$\delta \phi$---a variation of $\phi_{cl}$--- is a smooth function on $X$, and
$$L_0=-e^{-\phi_{cl}}\frac{\pa^2}{ \pa z \pa \bar{z}},$$
is a hyperbolic Laplacian acting on functions, i.e.~is Laplace-Beltrami
operator of the Poincar\'{e} metric on $X$. It is positively definite
self-adjoint operator in Hilbert space $H_0(X)$ of square integrable
functions on $X$ with respect to the volume form $d\rho$.

Second, ``mimic'' the saddle point method expansion in the finite-dimensional
case (we assume that the reader is familiar with it; see, e.g., \cite{LM} for
detailed exposition), replacing partial derivatives by variational derivatives
and matrices by Schwartz kernels (in the sense of distributions) of the
corresponding operators. There are two fundamental problems which one must
resolve along this way.

(a) One needs to define the determinant of differential operator $L_0+1/2$. If
it was an elliptic operator on a compact manifold, this could be done
in a standard fashion by using a heat kernel technique and zeta
function of the elliptic operator, or in physical terms, by using the proper
time regularization (see, e.g., \cite{Sw}). In our case Riemann surface $X$ is
not compact, operator $L_0$ has a $n$-fold continuous spectrum and a
point spectrum (see, e.g., \cite{Venkov}), so that the heat kernel approach
is not immediately applicable; additional regularization for the continuous
spectrum is needed. However, the determinants of hyperbolic Laplacians
can be also defined by means of the Selberg zeta function; for compact Riemann
surfaces this definition is equivalent to the standard heat-kernel
definition~\cite{FD,Sar}. Thus, following~\cite{ZTIT}, we set
$$\det(2L_0+1) \doteq Z_{X}(2),$$
where $Z_X(s)$ (see, e.g., \cite{Venkov}) is the Selberg zeta function of
Riemann surface $X$, defined by
$$Z_{X}(s) \doteq \prod_{m=0}^{\infty}\prod_{\{ l \}}(1-e^{-(s+m)|l|}).$$
Here $l$ runs over all simple closed geodesics on $X$ with respect to the
Poincar\'{e} metric with $|l|$ being the length of $l$; the infinite product
converges absolutely for ${\rm Re}s>1$.

(b) Recall the fundamental role played by the inverse matrix of the
Hessian of the classical action at the isolated critical point in the standard
finite-dimensional formulation of the steepest descent method~\cite{LM}. In our
infinite-dimensional case the analog of the Hessian is operator $2L_0+1$. Its
inverse $(2L_0+1)^{-1}$ is an integral operator, whose kernel---a propagator of
the theory---is given by Green's function $G(z, z^{\prime})$, which satisfies
on $X$ the following PDE
\beq \label{PDE}
-2G_{z \bar{z}}(z,z^{\prime}) + e^{\phi_{cl}(z)}G(z,z^{\prime})
=\delta(z-z^{\prime}).
\eeq

The Green's function can be uniquely characterized by the following properties.

{\bf 1} It is a smooth function on $X \times X \setminus D$, where $D$ is
diagonal $z^{\prime} =z$ in $X \times X$.

{\bf 2} It is symmetric: $G(z,z^{\prime})=G(z^{\prime},z),~z,z^{\prime} \in
X$.

{\bf 3} For fixed $z^{\prime} \in X$ it satisfies on $X \setminus \{z^{\prime}
\}$ the following PDE
$$(-2\pa^{2}_{z \bar{z}} +e^{\phi_{cl}(z)})G(z,z^{\prime})=0.$$

{\bf 4} For any $z \in X$ function
$$G(z,z^{\prime})+\frac{1}{2 \pi}\log|z-z^{\prime}|^2$$
is continuous as $z^{\prime} \rightarrow z$.

{\bf 5} For fixed $z^{\prime} \in X$ $G(z,z^{\prime}) \rightarrow 0$ as
$z \rightarrow z_i, ~i=1, \cdots, n$.

Now, the second problem one encounters in the functional case, is the problem
of short-distance divergences: Green's function $G$ and its partial derivatives
blow-up at coincident points. These divergences can be regularized in a
reparametrization invariant fashion by using the following asymptotic
expansion, which goes back to Hadamard~\cite{H}
\begin{eqnarray} \label{SDB}
G(z,z^{\prime}) = &-&\frac{1}{2 \pi}\{\log|z-z^{\prime}|^{2}(1+\frac{1}{2}
e^{(\phi_{cl}(z)+\phi_{cl}(z^{\prime}))/2}|z-z^{\prime}|^2 + \cdots)\}
\nonumber
\\
&-& \frac{1}{4 \pi}(\phi_{cl}(z) + \phi_{cl}(z^{\prime})) +\cdots,
\end{eqnarray}
as $z^{\prime} \rightarrow z$, where dots indicate higher order terms, which
can
be evaluated explicitly. Expansion (\ref{SDB}) admits
repeated differentiation with respect to $z$ and $z^{\prime}$. In particular,
it enables to regularize the logarithmic divergency of the Green's function on
the diagonal in a reparametrization invariant way
\beq \label{GZZ}
G(z,z) \doteq \lim_{z^{\prime} \rightarrow z}(G(z,z^{\prime})+
\frac{1}{2 \pi}(\log|z-z^{\prime}|^2 + \phi_{cl}(z))).
\eeq

This is a outline of the reparametrization invariant regularization scheme
for defining expectation value $<X>$.

It is instructive to present an equivalent description, which is based on the
Fuchsian uniformization of Riemann surface $X$.

First, recall the uniformization theorem (see, e.g.~\cite{Kra}), which states
that Riemann surface $X$ with $\chi(X)<0$ can be represented as a quotient of
hyperbolic
plane $H$---upper half-plane $H =\{\zeta \in \C ~|~ {\rm Im}\zeta > 0
\}$---by the fractional-linear action of a torsion-free finitely generated
Fuchsian group $\Gamma$
$$X \simeq H/\Gamma.$$
In other words, there exists a holomorphic covering $J: H \mapsto X$, with
$\Gamma$---a discrete subgroup of ${\rm PSL}(2,\R)$---acting as a group of
automorphisms (as an abstract group, $\Gamma$ is isomorphic to a fundamental
group of $X$). In terms of covering map $J$, classical solution $\phi_{cl}$
has the following explicit form
\beq \label{PHI-J}
e^{\phi_{cl}(z,\bar{z})}=\frac{|(J^{-1})^{\prime}(z)|^{2}}
{({\rm Im}~J^{-1}(z))^{2}},
\eeq
stating that Poincar\'{e} metric $e^{\phi_{cl}(z,\bar{z})}|dz|^{2}$ is a
projection on $X$ of hyperbolic metric $({\rm Im}\zeta)^{-2}|d\zeta|^{2} $
on $H$. This projection is well-defined, since the hyperbolic metric on $H$ is
${\rm PSL}(2,\R)$-invariant.

Second, hyperbolic Laplacian $L_0$, lifted to $H$, has the form
$$L_0=-({\rm Im}\zeta)^2 \frac{\pa^2}{\pa \zeta \pa \bar{\zeta}},$$
and Hilbert space $H_0(X)$ is isomorphic to the space of $\Gamma$-automorphic
functions,
$$f(\gamma\zeta)=f(\zeta),~\gamma \in \Gamma,$$
which are square integrable on $H/\Gamma$ with respect to the volume form
of the hyperbolic metric on $H$. Denote by ${\cal G}_{\Gamma}(\zeta,
\zeta^{\prime})$ the Green's function of operator $2L_0+1$ on $H/\Gamma$
and by ${\cal G}(\zeta, \zeta^{\prime})$---the Green's function of $2L_0+1$
on the upper half-plane $H$. It is well-known (see, e.g., \cite{Lang}), that
the Green's function ${\cal G}$ admits explicit representation
\beq \label{ER}
{\cal G}(\zeta,\zeta^{\prime})=\frac{1}{2 \pi}\int_{0}^{1}\frac{t(1-t)}{
(t+u)^2}dt= \frac{1}{2 \pi}(2u+1)\log(\frac{2u+1}{u})-\frac{1}{\pi},
\eeq
where
\beq \label{PPI}
u \doteq u(\zeta,\zeta^{\prime})=\frac{|\zeta-\zeta^{\prime}|^2}{4 {\rm Im}
\zeta~{\rm Im}\zeta^{\prime}},
\eeq
whereas Green's function ${\cal G}_{\Gamma}$ can be obtained by the method
of images
\beq \label{series}
{\cal G}_{\Gamma}(\zeta,\zeta^{\prime})=\sum_{\gamma \in \Gamma}{\cal G}
(\zeta,\gamma\zeta^{\prime}).
\eeq
Series in (\ref{series}) converges uniformly and absolutely on compact subsets
of $(H/\Gamma \times H/\Gamma)\setminus D$. It follows from (\ref{series}) that
the Green's function ${\cal G}_{\Gamma}$ is $\Gamma$-automorphic with respect
to both variables
\beq \label{AG}
{\cal G}_{\Gamma}(\gamma \zeta, \zeta^{\prime})={\cal G}_{\Gamma}(\zeta,
\gamma \zeta^{\prime})= {\cal G}_{\Gamma}(\zeta,\zeta^{\prime}),~\zeta \neq
\zeta^{\prime} \in H,~\gamma \in \Gamma.
\eeq
The Green's functions on $X$ and on $H/\Gamma$ are related by the following
simple formula
\beq \label{GFR}
G(z,z^{\prime})={\cal G}_{\Gamma}(J^{-1}(z),J^{-1}(z^{\prime})),~z \neq z^{
\prime} \in X,
\eeq
where according to (\ref{AG}), the right hand side does not depend on the
choice of a branch of multi-valued function $J^{-1}:X \mapsto H$. Using
representation (\ref{GFR}), properties of map $J$ (see, e.g., \cite{ZT1}),
and formula (\ref{PHI-J}), it is easy to see that regularization scheme
described above is essentially equivalent to the subtraction of contribution
${\cal G}(\zeta,\zeta^{\prime})$ of the unit element $I \in \Gamma$
from series (\ref{series}) as $z^{\prime} \rightarrow z$.
Thus, for instance,
\beq \label{GRD}
{\cal G}_{\Gamma}(\zeta,\zeta) \doteq \sum_{\gamma \in \Gamma, \gamma \neq I}
{\cal G}(\zeta,\gamma \zeta),
\eeq
and
\beq \label{GRDD}
\pa^{2}_{\zeta \zeta^{\prime}}{\cal G}_{\Gamma}|_{\zeta^{\prime}=
\zeta} \doteq \sum_{\gamma \in \Gamma, \gamma \neq  I} \pa^{2}_{\zeta
\zeta^{\prime}}{\cal G}(\zeta,\zeta).
\eeq
This prescription exhibits ``universal'' nature of the regularization
scheme, which takes full advantage of the knowledge of the underlying
hyperbolic geometry of Riemann surfaces---geometry of classical ``space-time''.

To summarize,  we have presented the set of rules for the perturbative
definition of expectation value $<X>$. In particular, at the tree level,
\beq \label{TREE}
\log <X>_{tree}=-\frac{1}{2 \pi h}S_{cl},
\eeq
and one-loop contribution is given by
\beq \label{LOOP}
{\cal X}_0 \doteq \log<X>_{loop}= -\frac{1}{2}\log \det(2L_0+1)=-\frac{1}{2}
\log Z_{X}(2).
\eeq
Higher multi-loop contributions can be also written down explicitly.
However, in their definition additional regularization of integrals
$$\int_{X}G^k(z,z)d\rho$$
at the punctures is required. This regularization is similar to the one used
in the derivation of the Selberg trace formula (see, e.g., \cite{Venkov}).

\subsection{Generating Functional $Z(\mu, \bar{\mu};X)$}

Here we derive perturbation expansion for the generating functional for
correlation functions involving stress-energy tensor components. To
illustrate our approach, we consider first the example of free bosons on $\C$,
where one can perform all calculations explicitly.

Namely, let $\psi$ be a real-valued scalar field on $\C$ with the classical
action given by Dirichlet functional
$$S_{free}(\psi) \doteq \int|\psi_z|^2 d^2z,$$
which defines a conformal theory of free bosons with stress-energy tensor
$$T_{free}(\psi)=-\psi_{z}^2/2h,~\bar{T}_{free}(\psi)=-\psi_{\bar{z}}^{2}/2h.$$
Generating functional for the correlation functions of these stress-energy
tensor components is given by the following functional integral
\beq \label{FFI}
Z_{free}(J,\bar{J}) \doteq \int{\cal D}\psi \exp\{-\frac{1}{2 \pi
h}S_{free}(\psi)+\int(T_{free}(\psi)J+\bar{T}_{free}(\psi)\bar{J})d^2z\},
\eeq
where $J$ is an external source. Functional integral (\ref{FFI}) is
Gaussian and its integrand is an exponential of a quadratic form of the
following differential operator
$$-\frac{1}{2 \pi h}(L_{free}-\pi \pa_z J \pa_z - \pi \pa_{\bar{z}} \bar{J}
\pa_{\bar{z}}),$$
where $L_{free}=-\pa^{2}_{z \bar{z}}$ is a Laplacian of Euclidean metric
$|dz|^2$ on $\C$. Denoting $G_{free}=(L_{free})^{-1}$, we get as a result of
Gaussian integration
\beq \label{GI}
{\cal Z}_{free}(J,\bar{J}) \doteq Z_{free}(J,\bar{J})/Z_{free}(J,\bar{J})|_{J=
\bar{J}=0}=\det(1-\pi G_{free}(\pa J \pa +\bar{\pa}\bar{J} \bar{\pa}))^{-1/2}.
\eeq
Therefore
\begin{eqnarray} \label{WFREE}
\frac{1}{h}{\cal W}_{free}(J,\bar{J})&=&-\frac{1}{2}\log \det(1-\pi G_{free}
(\pa J \pa + \bar{\pa} \bar{J} \bar{\pa})) \nonumber \\
&=& \sum_{k=1}^{\infty}\frac{\pi^k}{2k}{\rm tr}
\{G_{free}(\pa J \pa + \bar{\pa} \bar{J} \bar{\pa})\}^k,
\end{eqnarray}
where the sum contains only the terms with even $k$. In particular, one gets
from (\ref{WFREE}) that
$$<<T_{free}(z)T_{free}(w)>> \doteq \frac{1}{h}\frac{\delta^2 {\cal W}_{free}}
{\delta J(z) \delta J(w)}|_{J=\bar{J}=0}=\frac{1/2}{(z-w)^4},$$
so that $c=1$, as it should be for the theory of free bosons.

Next, consider the case of the Liouville theory. We will describe the
perturbation expansion of generating functional (\ref{Z}).

First, write $\phi=\phi_{cl} + \chi$ and expand the Liouville action and
the stress-energy tensor around classical solution $\phi_{cl}$. Using
(\ref{LAE}) with $\delta \phi$ replaced by $\chi$, we get
\beq \label{LAE2}
S(\phi_{cl} + \chi)=S_{cl} +\frac{1}{2}\int_{X} \chi(2L_0+1)(\chi) d\rho
+S_{int}(\chi),
\eeq
where
\beq \label{INT}
S_{int}(\chi)=\sum_{k=3}^{\infty}\frac{1}{k!}\int_{X}\chi^k d\rho,
\eeq
and
\beq \label{TE}
T(\phi_{cl}+\chi)=T_{cl}+\frac{1}{h}(\chi_{zz}-\chi_{z}(\phi_{cl})_z
-\frac{1}{2}\chi^{2}_{z}).
\eeq

Second, using Stokes' formula, we obtain
$$\int_{X}T(\phi) \mu d^2z=\int_{X}T_{cl}\mu d^2z+\frac{1}{h}\int_{X}
(-\frac{1}{2}\chi_{z}^{2}+\chi \omega)d^2z,$$
where
\beq \label{ommu}
\omega=\mu_{zz}+(\phi_{cl})_z\mu_z+(\phi_{cl})_{zz}\mu,
\eeq
as it follows from from transformation laws (\ref{ConF}) and (\ref{BD}), is
well-defined $(1,1)$-form on $X$.

Third, using the variational analog of identity
$$F(x)e^{\lambda x}=F(\frac{d}{d \lambda}) e^{\lambda x},$$
and expansions (\ref{LAE2})---(\ref{TE}), generating functional (\ref{Z})
can be represented as classical term
$$\exp\{ {\rm v.p.}\int_{X}(T_{cl} \mu+ \bar{T}_{cl} \bar{\mu})d^2z\}$$
times the result of an application of the following ``pseudo-variational''
operator
$${\cal S}_{int}(h\frac{\delta}{\delta \xi}) \doteq \exp\{-\frac{1}{2 \pi h}
S_{int}(h\frac{\delta}{\delta \xi})\}$$
to Gaussian functional integral
$$
\int_{{\cal C}(X)}{\cal D}\chi \exp\{-\frac{1}{2 \pi h}\int_{X}\{\chi(L_0+1/2)
(\chi)+ \pi(\chi^{2}_{z} \mu  + \chi^{2}_{\bar{z}} \bar{\mu} -\xi \chi)\}d \rho
\},
$$
where
\beq \label{xif}
\xi=f+\bar{f},~f=e^{-\phi_{cl}}\omega,
\eeq
and $f$ is a globally defined function on $X$. Gaussian integration with
respect to $\chi$ can be performed explicitly. The final result reads
\begin{eqnarray} \label{ZFinal}
Z(\mu, \bar{\mu};X) &=&\exp\{{\rm v.p.}\int_{X}(T_{cl}\mu+\bar{T}_{cl}
\bar{\mu})d^{2}z\}\det\{\frac{G}{(1-2\pi G e^{-\phi_{cl}}(\pa \mu \pa+
\bar{\pa} \bar{\mu} \bar{\pa}))}\}^{1/2} \nonumber \\
& & \mbox{}{\cal S}_{int}(h \frac{\delta}{\delta \xi})
(\exp\{\frac{\pi}{h}\int_{X}\xi G(1-2 \pi Ge^{-\phi_{cl}} (\pa \mu
\pa + \bar{\pa}\bar{\mu} \bar{\pa}))^{-1} (\xi) d\rho \}),
\end{eqnarray}
where $G=(2L_0 +1)^{-1}$. It is understood that one should first apply
${\cal S}_{int}(\delta/\delta \xi)$) with subsequent substitution $\xi=
e^{-\phi_{cl}}(\omega+\bar{\omega})$.

Note that determinant contribution to this formula looks similar to free theory
case (\ref{GI}) after replacement $\phi_{cl} \mapsto 0$ and $G \mapsto
G_{free}$. Specific features of the Liouville theory are reflected in the
second line of (\ref{ZFinal}).

{}From (\ref{ZFinal}) it is straightforward to get perturbation expansion for
generating functional ${\cal W}$. As in the free case, one should apply
standard rule
$$\log \det(1-A)=\sum_{k=1}^{\infty}\frac{1}{k}{\rm tr}A^k,$$
expand the exponentials in (\ref{ZFinal}) and use the regularization scheme,
indicated in \S3.1.

\subsection{Examples}

Here we illustrate our formalism by considering perturbation expansion
of one - and two-point correlation functions.

{\bf (i)} {\it One-point correlation function $<<T(z)X>>$}

According to definition (\ref{CCF}),
\beq \label{ST}
<<T(z)X>>=\frac{1}{h}\frac{\delta {\cal W}}{\delta \mu}|_{\mu=0},
\eeq
and it is clear that
\beq \label{TTREE}
<<T(z)X>>_{tree}=T_{cl}(z).
\eeq

In order to obtain the expression for $<<T(z)X>>_{loop}$, we need to collect
all
terms of order $h^0$ in formula (\ref{ST}). One can get these terms in
two different ways.

First, consider such terms as coming from the ``universal'' expression $\log
\det$. It always has order $h^0$ and contributes to the one-loop expansion of
$<<T(z)X>>$ through
$$\frac{\delta{\rm tr}(G\pa \mu \pa)}{\delta \mu(z)}|_{\mu=0}=
G_{z z^{\prime}}|_{z^{\prime}=z}.$$

Second, terms of order $h^0$ come from the interaction. Namely, expand
$\log{\cal S}_{int}(h\delta/\delta \xi)$ into power series in $h$ and apply
its first term (which is of order $h^2$) to the first term of the corresponding
expansion of
$$\exp\{\frac{\pi}{h} \int_{X}\xi(G(1-2 \pi G e^{-\phi_{cl}}( \pa \mu \pa +
\bar{\pa} \bar{\mu} \bar {\pa}))^{-1}(\xi)d\rho\},$$
keeping only terms quadratic in $\xi$; according to (\ref{ommu}), they
also are quadratic in $\mu$. The result is proportional to
\beq \label{int}
\frac{\delta}{\delta \mu(z)}\{ \int_{X}G(z^{\prime},z^{\prime})
G(\xi)(z^{\prime})d\rho^{\prime}\}=\int_{X} G(z^{\prime},z^{\prime}){\cal D}_z
(G)(z,z^{\prime})d\rho^{\prime}.
\eeq
Here, ${\cal D}_z \doteq \pa^{2}_{zz}-(\phi_{cl})_z \pa_z$, and
we have used (\ref{ommu})---(\ref{xif}) and the Stokes' formula.

Thus we obtain
\beq \label{1T}
<<T(z)X>>_{loop}=-\pi G_{z z^{\prime}}|_{z^{\prime}=z}-{\rm v.p.}\pi\int_{X}
G(z^{\prime},z^{\prime}){\cal D}_z(G)(z,z^{\prime})d\rho^{\prime}.
\eeq
Here $G(z,z)$ is given by (\ref{GZZ}) and in accordance with (\ref{SDB}), the
regularized value of  $G_{zz^{\prime}}$ at diagonal $z^{\prime}=z$ is
defined as
\beq \label{GPZZ}
G_{zz^{\prime}}|_{z^{\prime}=z} \doteq \lim_{z^{\prime} \rightarrow z}(
G_{zz^{\prime}}(z,z^{\prime})+\frac{1}{2 \pi}(\frac{1}{(z-z^{\prime})^2}
-\frac{1}{2}e^{\phi_{cl}(z)}\frac{\bar{z}-\bar{z}^{\prime}}{z-z^{\prime}})).
\eeq
Note that the integral in (\ref{1T}) is a principal value integral, since
${\cal D}_z(G) (z,z^{\prime})$ has a second order pole at $z^{\prime}=z$.

{\bf (ii)} {\it Two-point correlation functions $<<T(z)T(w)X>>$ and
$<<T(z)\bar{T}(\bar{w})X>>$}

According to definition (\ref{CCF}),
\beq \label{TTW}
<<T(z)T(w)X>>=\frac{1}{h}\frac{\delta^2 {\cal W}}{\delta \mu(z) \delta
\mu(w)}|_{\mu=0}.
\eeq
At the tree level, this correlation function has order $h^{-1}$ and the only
term of this order comes from integral (\ref{int}) and is quadratic
in $\xi$. We get the following expression
\beq \label{treeTT}
<<T(z)T(w)X>>_{tree}=\frac{2 \pi}{h}{\cal D}_z{\cal D}_wG(z,w).
\eeq

Similarly,
\beq \label{treeTCT}
<<T(z)\bar{T}(\bar{w})X>>_{tree}=\frac{2 \pi}{h} {\cal D}_z {\cal D}_{\bar{w}}
G(z,w).
\eeq

One-loop contributions to $<<T(z)T(w)X>>$ and $<<T(z)\bar{T}(\bar{w})X>>$
are more complicated and consist of several terms. In particular, $<<T(z)T(w)X
>>_{loop}$ contains term
$$2 \pi^2 G_{zw}^{2}(z,w),$$
which comes from expansion of the universal expression $\log \det$.
As we shall see in \S4.2, it is this term that contributes to the quantum
correction to the central charge.

{\bf Problems}

{\bf 1} Calculate two-loop contribution ${\cal X}_1$ to $\log <X>$.

{\bf 2} Show the equivalence of regularization schemes on $X$ and on
$H/\Gamma$. In particular, prove that
$$G(J(\zeta),J(\zeta))={\cal G}_{\Gamma}|_{\zeta^{\prime}=\zeta}~-~\frac{1-
\log 2}{\pi}$$
and
$$G_{zz^{\prime}}|_{z^{\prime}=z=J(\zeta)}~J^{\prime}(\zeta)^2=\pa^{2}_{\zeta
\zeta^{\prime}}{\cal G}_ {\Gamma}|_{\zeta^{\prime}=\zeta}~+~ \frac{1}{12 \pi}
{\cal S}(J)(\zeta).$$

{\bf 3} Complete all details of the calculations of generating functional
$Z_{free}(J,\bar{J})$.

{\bf 4} Complete details for the Liouville case; in particular, verify
formulas (\ref{1T})---(\ref{treeTCT}).

{\bf 5} Compute two-loop contribution to correlation function $<<T(z)X>>$
and one-loop contribution to $<<T(z)T(w)X>>$ and $<<T(z)\bar{T}(\bar{w})X>>$.

\section{Ward Identities and Modular Geometry}

\subsection{Semi-classical Approximation}

{\bf (i)} {\it One-point correlation function $<<T(z)X>>$}

Consider Ward identity (\ref{NT}). At the tree level it reads
\beq \label{treeNT}
T_{cl}(z)-T_{s}(z)=-\frac{1}{2 \pi h}{\cal L}(z)S_{cl}.
\eeq
{}From (\ref{L})---(\ref{R}) it follows that the right hand side of
(\ref{treeNT}) has only simple poles at $z=z_i$. According to (\ref{CSET}) and
(\ref{TS}), it is equivalent to the statement that classical dimension of the
puncture operator is $1/2h$ (cf.~discussion in \S1.1 and \S2.1). Using
constraints (\ref{C}), we then have
$$T_{cl}(z)-T_{s}(z)=\frac{1}{h} \sum_{i=1}^{n-3}c_iR(z,z_i),$$
so that (\ref{treeNT}) is equivalent to the following relations
\beq \label{AP}
c_i=-\frac{1}{2 \pi} \pa_i S_{cl},~i=1,\cdots, n-3.
\eeq
In other words, coefficients $c_i$ are ``conjugate'' to punctures $z_i$, and
classical action $S_{cl}$ plays a role of generating function.

What is the meaning of coefficients $c_i$? Actually, they have been around for
more than hundred years, were called ``accessory parameters'' by Poincar\'{e}
and are closely related with the Fuchsian uniformization of Riemann surface
$X$. Namely, as it follows from (\ref{PHI-J}),
\beq \label{TSW}
T_{cl}(z)=\frac{1}{h}{\cal S}(J^{-1})(z).
\eeq
Next, according to Klein and Poincar\'{e} (see, e.g.~\cite{T3} for more
details), the Schwarzian derivative of the inverse function to uniformization
map $J$ has the property that the monodromy group of associated second order
linear differential equation
$$\frac{d^2y}{dz^2}+\frac{1}{2}{\cal S}(J^{-1})(z)y=0,~z \in X,$$
coincides (up to a conjugation) with Fuchsian group $\Gamma$, uniformizing
Riemann surface $X$! Therefore, in the representation
\beq \label{SWC}
{\cal S}(J^{-1})(z)=\sum_{l=1}^{n-1}(\frac{1}{2(z-z_i)^2}+\frac{c_i}{z-z_i}),
\eeq
which follows from (\ref{CSET}) and (\ref{TSW}), coefficients $c_i$
are uniquely determined by this global condition. It turns out to be
an extremely difficult mathematical problem to characterize accessory
parameters as functions of the punctures, i.e.~as functions on $Z_n$. On the
other hand, CWI (\ref{treeNT}) implies that coefficients $c_i$ possess
a remarkable property (\ref{AP})! These relations, conjectured by
Polyakov~\cite{Pol2} on the basis of CWI, were rigorously proved in~\cite{ZT1}.
The proof is based on fundamental facts from the Teichm\"{u}ller theory, which
we present in \S4.3.

{\bf (ii)} {\it Two-point correlation functions $<<T(z)T(w)X>>$ and
$<<T(z)\bar{T}(\bar{w})X>>$}

Start with the CWI (\ref{NTCT}). Using (\ref{treeTCT}), at the tree
level it can be written as
\beq \label{treeNTCT}
4{\cal D}_z {\cal D}_{\bar{w}}G(z,w)=-\frac{1}{\pi^2}{\cal L}(z)\bar
{{\cal L}}(\bar{w})S_{cl}.
\eeq
Denoting $P(z,w) \doteq 4{\cal D}_z {\cal D}_{\bar{w}}G(u,v)$, we can rewrite
(\ref{treeNTCT}) as the following relation on $Z_n$
\beq \label{WPP}
P(z,w)dz^2d\bar{w}^2=-\pa \bar{\pa}S_{cl}.
\eeq

It turns out that kernel $P(z,w)$ can be described explicitly: it is
finite-dimensional and is related with the Weil-Petersson Hermitian metric on
space $Z_n$.
Namely, let $H_2(X)$ be the Hilbert space of quadratic differentials on $X$---
tensors of type $(2,0)$, which are square integrable with respect to
the measure $e^{-\phi_{cl}}d^2z$ on $X$. If $P, Q \in H_2(X)$, then
\beq \label{IP}
(P,Q) \doteq \int_{X}P(z)\overline{Q(z)}e^{-\phi_{cl}(z)}d^2z < \infty.
\eeq
Denote by ${\cal H}^{2,0}(X)$ the space of harmonic quadratic differentials
on $X$, i.e.~the subspace in $H_2(X)$, consisting of holomorphic quadratic
differentials---zero modes of operator $\bar{\pa}$ in the space $H_2(X)$, that
is a finite-dimensional vector space of complex dimension $n-3$. It is easy to
show that $P(z,w)$ is the kernel of orthogonal Hodge projection operator
$$P:H_2(X) \mapsto {\cal H}^{2,0}(X).$$

Indeed, using formulas (\ref{PHI-J})---(\ref{series}), one gets
\beq \label{PCP}
P(z,w)={\cal P}_{\Gamma}(J^{-1}(z),J^{-1}(w))(J^{-1})^{\prime}(z)^2
(\bar{J}^{-1})^{\prime}(\bar{w})^2,
\eeq
where
\beq \label{CP}
{\cal P}_{\Gamma}(\zeta,\zeta^{\prime})=\frac{12}{\pi} \sum_{\gamma \in
\Gamma}\frac{\gamma^{\prime}( \overline{\zeta^{\prime}})^2}
{(\zeta-\gamma \overline{\zeta^{\prime}})^4}.
\eeq
Next, recall that Hilbert space $H_2(X)$, lifted to upper half-plane
$H$ by map $Q \mapsto q=Q \circ J (J^{\prime})^2$, is isomorphic to
space $H_2(H/\Gamma)$, consisting of automorphic forms of weight $4$ with
respect to Fuchsian group $\Gamma$,
$$q(\gamma \zeta)\gamma^{\prime} (\zeta)^2=q(\zeta),~\gamma \in \Gamma,$$
which are square integrable on $H/\Gamma$ with respect to measure
$({\rm Im}\zeta)^2 d^2\zeta$. Therefore, for any $q \in H_2(H/\Gamma)$,
$${\cal P}_{\Gamma}(q)=\int_{H/\Gamma}{\cal P}_{\Gamma}(\zeta,\zeta^{\prime})
q(\zeta^{\prime})({\rm Im}\zeta^{\prime})^2 d^2 \zeta^{\prime}=\frac{12}{\pi}
\int_{H}\frac{({\rm Im}\zeta^{\prime})^2}{(\zeta-\overline{\zeta^{\prime}})^4}
q(\zeta^{\prime})d^2\zeta^{\prime}.$$
According to the Ahlfors lemma~\cite[Ch.~VI.D, Lemma 2]{A}, the
result of integration is either $q$, if $q \in {\cal H}^{2,0}(H/\Gamma)$, or
zero, if $q$ is orthogonal to ${\cal H}^{2,0}(H/\Gamma)$.

In terms of arbitrary basis $P_i$ in ${\cal H}^{2,0}(X)$, the kernel $P(z,w)
$ can be written as
\beq \label{PP}
P(z,w)=\sum_{i,j=1}^{n-3}g^{ji}P_{i}(z)\overline{P_{j}(w)},
\eeq
where matrix $\{g^{ij}\}$ is inverse to the Gram matrix of the basis
$P_i$ with respect to inner product (\ref{IP}), i.e.~the matrix with
elements $g_{ij}=(P_i,P_j),~i,j=1, \ldots,n-3$. In particular, if $P_i$ is a
basic, dual to the basic of harmonic Beltrami differentials corresponding to
vector fields $\pa/\pa z_i$, then
\beq \label{WP1}
g^{ij}=(\frac{\pa}{\pa z_i},\frac{\pa}{\pa z_j})_{WP},
\eeq
where $(~,~)_{WP}$ stands for the Weil-Petersson Hermitian metric on $Z_n$ (see
\S4.3). Combining formulas (\ref{WPP}), (\ref{PP}) and (\ref{WP1}), we
finally obtain
\beq \label{WPP1}
(\frac{\pa}{\pa z_i},\frac{\pa}{\pa z_j})_{WP}=-\frac{\pa^2 S_{cl}}{\pa z_i \pa
\bar{z}_j}.
\eeq

Relation (\ref{WPP1}), which is equivalent to CWI (\ref{NTCT}) at the tree
level, states that the Weil-Petersson metric on space $Z_n$ is K\"{a}hler,
with a potential given by classical action $S_{cl}$! This potential is a
smooth single-valued function on $Z_n$, but not on moduli space
${\cal M}_{0,n}$, where it is a Hermitian metric in a certain holomorphic
line bundle over ${\cal M}_{0,n}$ (see~\cite{Zog}). The K\"ahler property of
the Weil-Petersson metric on the Teichm\"{u}ller space was proved by Ahlfors
and Weil~\cite{W,A2}. Relation (\ref{WPP1}), which establishes a connection
between Fuchsian uniformization
and Weil-Petersson geometry through classical Liouville action $S_{cl}$ as
the K\"{a}hler potential, was not known to the founders of the Teichm\"{u}ller
theory. It was proved only recently~\cite{ZT1}, after the importance of the
quantum Liouville theory~\cite{Pol1,Pol2} has been appreciated. Comparing
(\ref{AP}) and (\ref{WPP1}) yields
\beq \label{APWP}
\frac{\pa c_i}{\pa \bar{z}_j}=\frac{1}{2 \pi}(\frac{\pa}{\pa z_i},\frac{\pa}
{\pa z_j})_{WP},
\eeq
so that the Weil-Petersson metric also ``measures'' a deviation of accessory
parameters $c_i$ from being holomorphic functions on $Z_n$.

Finally, consider CWI (\ref{NTT}). Using (\ref{TTREE}) and (\ref{treeTT}),
we get at the tree level
\beq \label{EI}
\frac{2 \pi}{h}{\cal D}_z{\cal D}_wG(z,w)=\frac{c_{cl}/2}{(z-w)^4}+
(2R_w(z,w)+R(z,w)\frac{\pa}{\pa w}+ {\cal L}(z))T_{cl}(w).
\eeq
Similarly to the previous case, denoting $\tilde{P}(z,w) \doteq 4{\cal D}_z
{\cal D}_w G(z,w)$ and setting
$$\tilde{{\cal P}}_{\Gamma}(\zeta,\zeta^{\prime})=\tilde{P}(J(\zeta),J(\zeta^
{\prime}))J^{\prime}(\zeta)^2J^{\prime}(\zeta^{\prime})^2,$$
we get
\beq \label{TilP}
\tilde{{\cal P}}_{\Gamma}(\zeta,\zeta^{\prime})={\cal P}(\zeta,\bar{\zeta}^
{\prime})= \frac{12}{\pi}\sum_{\gamma \in \Gamma}\frac{\gamma^ {\prime}(\zeta^
{\prime})^2}{(\zeta-\gamma \zeta^{\prime})^4}.
\eeq
This symmetric kernel $\tilde{{\cal P}}_{\Gamma}(\zeta,\zeta^{\prime})=
\tilde{{\cal P}}_{\Gamma}(\zeta^{\prime},\zeta)$ is the third derivative of
the so-called meromorphic Eichler integral for Fuchsian group $\Gamma$,
which is regular at the punctures and has a simple pole at $\zeta=\zeta^
{\prime}$ (see~\cite[Ch.V, Sect.~7]{Kra}), thus establishing the connection
between CWI (\ref{EI}) and Eichler integrals. The right hand side of
(\ref{EI}), in virtue of (\ref{treeNT}) and (\ref{AP}), essentially consists
of $\pa c_i/\pa z_j$, partial derivatives of accessory parameters in
holomorphic directions. In~\cite{ZT1} we proved explicit formulas expressing
these derivatives through kernel $\tilde{P}(z,w)$~\cite[\S4.3]{ZT1}; it is
easy to show that these results can be put together into single relation
(\ref{EI})!
Moreover, as it follows from (\ref{treeTT}) and (\ref{TilP}),
$$<<T(z)T(w)X>>=\frac{6/h}{(z-w)^4}+O(|z-w|^{-3})~~{\rm as}~z \rightarrow w,$$
so that, as expected, $c_{cl}=12/h$.

Thus we have seen that even at the tree level, CWI for the two-dimensional
quantum gravity are highly non-trivial and yield new important information
about the uniformization of the Riemann surfaces and complex geometry of moduli
spaces.

\subsection{One-loop Approximation}

{\bf (i)} {\it One-point correlation function $<<T(z)X>>$}

Consider CWI (\ref{NT}) at the one-loop level: using formulas (\ref{LOOP}) and
(\ref{1T}) for one-loop contributions to the $\log<X>$ and $<<T(z)X>>$, we
get
\beq \label{WI1T}
2\pi G_{zz^{\prime}}|_{z^{\prime}=z}+2\pi {\rm v.p.}\int_{X}G(z^{\prime},
z^{\prime}){\cal D}_zG(z,z^{\prime})d\rho^{\prime}={\cal L}(z)\log Z_{X}(2).
\eeq

First, it is not obvious that the left hand side of (\ref{WI1T}) is a
holomorphic quadratic differential on $X$, as it is explicitly stated in
the right hand side. However, consider the following formal calculations
\begin{eqnarray} \label{1FC}
\frac{\pa}{\pa \bar{z}}(G_{zz^{\prime}}|_{z^{\prime}=z}) & = & ((\pa_{\bar{z}}+
\pa_{\bar{z}^{\prime}})G_{zz^{\prime}})|_{z^{\prime}=z} \nonumber \\
&=& (G_{z \bar{z} z^{\prime}}(z,z^{\prime})+G_{zz^{\prime}\bar{z}^{\prime}}
(z,z^{\prime}))|_{z^{\prime}=z} \nonumber \\
&=& \frac{1}{2}(e^{\phi_{cl}(z)}G_{z^{\prime}}(z,z^{\prime})+
e^{\phi_{cl}(z^{\prime})} G_{z}(z,z^{\prime})
-\delta_{z^{\prime}}(z-z^{\prime})-\delta_{z}(z-z^{\prime}))|_{z^{\prime}=z}
\nonumber \\
&=& \frac{1}{2}e^{\phi_{cl}}(z)\frac{d}{dz}G(z,z),
\end{eqnarray}
and
\begin{eqnarray} \label{2FC}
\pa_{\bar{z}}(G_{zz}-(\phi_{cl})_zG_{z})&=&G_{zz \bar{z}}-(\phi_{cl})_
{z\bar{z}}G_{z} -(\phi_{cl})_zG_{z \bar{z}} \nonumber \\
&=&\frac{1}{2}(e^{\phi_{cl}}G_{z} + (\phi_{cl})_z e^{\phi_{cl}}G -e^{\phi_{cl}}
\delta_{z}-(\phi_{cl})_ze^{\phi_{cl}}\delta-
e^{\phi_{cl}}G_{z} \nonumber \\
& & - (\phi_{cl})_ze^{\phi_{cl}}G + (\phi_{cl})_ze^{\phi_{cl}}
\delta)=-\frac{1}{2}e^{\phi_{cl}}\delta_{z},
\end{eqnarray}
which use equations (\ref{LE}), (\ref{PDE}) and symmetry property {\bf 2} of
the Green's function. Formula (\ref{2FC}) implies that
\beq \label{GG}
\frac{\pa}{\pa \bar{z}}(\int_{X}G(z^{\prime},z^{\prime}){\cal D}_zG(z,z^
{\prime})d\rho^{\prime}) =-\frac{1}{2}e^{\phi_{cl}(z)}\frac{d}{dz}G(z,z),
\eeq
which, together with (\ref{1FC}), ``show'' that $<<T(z)X>>_{loop}$ is
holomorphic on $X$.

One can easily make these arguments rigorous (and valid for the case of compact
Riemann surfaces as well).

First, observe that expression (\ref{1T}) has a transformation law of a
projective connection (times $1/6$). Indeed, from (\ref{GZZ}) and classical
formula
$$\lim_{z^{\prime} \rightarrow z}\{\frac{f^{\prime}(z)f^{\prime}(z^{\prime})}{
(f(z)-f(z^{\prime}))^{2}}- \frac{1}{(z-z^{\prime})^{2}}\}
=\frac{1}{6}{\cal S}(f)(z),$$
where $f$ is a (locally) holomorphic function (see, e.g., \cite{Tu}), it
follows that the first term in (\ref{1T}) has transformation law (\ref{PC})
(times $1/6$). Moreover, as it follows from (\ref{ConF}), the differential
operator ${\cal D}$ maps functions on $X$ into quadratic differentials.
Therefore, the second term in (\ref{1T}) is a well-defined quadratic
differential on $X$.

Second, using definition (\ref{GPZZ}) of $G_{zz^{\prime}}|_{z^{\prime}=z}$,
property {\bf 3} of the Green's function, asymptotics
$$G_{z}=-\frac{1}{2 \pi}\{\frac{1}{z-z^{\prime}} + \frac{1}{2}
(\phi_{cl})_z\} + O(1),~z^{\prime} \rightarrow z,$$
and the Taylor formula, one gets
$$\frac{\pa}{\pa \bar{z}}(G_{zz^{\prime}}|_{z^{\prime}=z})=
\frac{1}{2}e^{\phi_{cl}(z)}\frac{d}{dz}G(z,z).$$
This justifies (\ref{1FC}).

Third, the same property {\bf 3} and equation (\ref{LE}) imply that
${\cal D}_zG(z,z^{\prime})$ is a holomorphic quadratic differential on
$X \setminus\{z^{\prime}\}$. It has the following asymptotics
$${\cal D}_zG(z,z^{\prime})=\frac{1}{2 \pi}\{\frac{1}{(z-z^{\prime})^{2}}+
\frac{(\phi_{cl})_{z^{\prime}}}{z-z^{\prime}}\} + O(1),~z^{\prime} \rightarrow
z,$$
so that (\ref{GG}) follows from the standard properties of singular
integrals~\cite[Ch.~5.A, Lemma 2]{A}.

Next, what about the validity of CWI (\ref{WI1T}) itself? First, it is not
difficult to show that
\beq \label{TCusp}
<<T(z)X>>_{loop}=P(-\pi G_{zz^{\prime}}|_{z^{\prime}=z}),
\eeq
where $P$ is the Hodge projector, so that $<<T(z)X>>_{loop} \in {\cal H}^{2,0}
(X)$. In particular, it has no second order poles at the punctures (see \S4.3),
so that at one-loop there is no contribution to the conformal dimension, in
agreement with the discussion in \S2.1. Second, consider explicit formula for
the first variation of the Selberg zeta function with respect to moduli, proved
in~\cite[Lemma 3]{ZTIT}, which states (after setting $s=2$) that
\beq \label{VZ}
\pa \log Z(2)=-2P(G_{zz^{\prime}}|_{z^{\prime}=z}).
\eeq
Combining (\ref{TCusp}) and (\ref{VZ}), we get (\ref{WI1T}),
which ``encodes'' a variational formula for the Selberg zeta function at a
special point $s=2$!

{\bf (ii)} {\it Two-point correlation functions $<<T(z)T(w)X>>$ and
$<<T(z)\bar{T}(\bar{w})X>>$}

Without presenting explicit (though quite complicated) expressions for
correlation function $<<T(z)T(w)X>>_{loop}$, here we restrict ourselves to a
simpler problem of determining its leading singularity as $z \rightarrow w$.
Direct inspection shows that this singularity is of order
$(z-w)^{-4}$ and can be obtained only from term $2 \pi^2G_{zw}^{2}(z,w)$,
described in \S3.2. According to property {\bf 4} of the Green's function, it
has the form
$$\frac{1/2}{(z-w)^4},$$
which shows that one-loop correction $c_{cl}$ to the central charge equals
$1$. Analyzing higher-loop terms, that can only be obtained from the
``interaction part'' in the expression for generating function $Z(\mu,
\bar{\mu};X)$, one can convince oneself that they do not contribute to the
leading singularity of $<<T(z)T(w)X>>$. Thus we have
$$c=c_{cl}+c_{loop}=\frac{12}{h}+1$$
for the central charge of the Virasoro algebra. This is in accordance with
general ``philosophy'' of the quantum field theory that only one-loop
approximation contributes to the anomaly.

It is also possible to analyze the CWI (\ref{NTCT}) at the one-loop level.
Working out explicit expression for $<<T(z)\bar{T} (\bar{w})X>>_{loop}$, one
can show that (\ref{NTCT}) in a one-loop approximation is equivalent to the
local index theorem for families of $\bar{\pa}$-operators on punctured Riemann
surfaces (a generalization of the Belavin-Knizhnik theorem~\cite{BK}), proved
in~\cite{ZTIT}.

\subsection{Elements of the Teichm\"{u}ller Theory}

Here we present, in a succinct form, basic facts from the Teichm\"{u}ller
theory
as it was developed by Ahlfors and Bers~\cite{A,AB,B}; our exposition mainly
follows~\cite{ZT1}. Although we restrict ourselves to the case of punctured
spheres, compact Riemann surfaces are treated similarly.

Recall that a Riemann surface $X$ is called marked, if a particular
canonical system of generators (up to an inner automorphism) of its fundamental
group is chosen. Two marked Riemann surfaces are isomorphic if there exists
a complex analytic isomorphism that maps one set of generators into the other
(up to a overall conjugation). This equivalence relation is more restrictive
than the usual one, so that the corresponding quotient space---Teichm\"{u}ller
space ${\cal T}_{0,n}$ of Riemann surfaces of genus zero with $n$
punctures---covers the moduli space ${\cal M}_{0,n}$ (and is easier to deal
with, since it is isomorphic to an open cell in $\C^{n-3}$).  All points in
${\cal T}_{0,n}$---equivalence classes of marked Riemann surfaces---can be
obtained from a given point---equivalence class of a marked Riemann surface
$X$---by deformations of its complex structure.

These deformations can be described by considering the hyperbolic plane $H$
first. Complex structures on $H$ correspond to the $1$-forms $d\zeta+ \mu d\bar
{\zeta}$, where $\mu \in L^{\infty}(H)$ is a $(-1,1)$-form on $H$ with the
property $\|| \mu \||_{\infty}<1$. The
standard complex structure on $H$ corresponds to the case $\mu=0$. According
to the fundamental theorem from the theory of quasi-conformal mappings, all
complex structures on $H$ are isomorphic: there exists diffeomorphism $f:H
\mapsto H$ satisfying Beltrami equation
$$\frac{\pa f}{\pa \bar{\zeta}}= \mu \frac{\pa f}{\pa \zeta}$$
on $H$. However, complex structure $d\zeta+\mu d \bar{\zeta}$ on $H$ can be
projected on $X \cong H/\Gamma$ only when $\mu$ is a Beltrami differential on
$H/\Gamma$, i.e.~satisfies the transformation law
$$\mu(\gamma \zeta) \frac{\overline{\gamma^{\prime}(\zeta)}}{\gamma^{\prime}
(\zeta)}= \mu (\zeta), ~\gamma \in \Gamma.$$
In this case, $\mu$ can be projected on $X$, so that
$M \doteq m \circ J^{-1}\overline{J^{-1^{\prime}}}/J^{-1^{\prime}}$
is a Beltrami differential on Riemann surface $X$. Even in this case
isomorphism $f$ will not necessarily be $\Gamma$-equivariant: Fuchsian group
$\Gamma^{\mu}=f \circ \Gamma \circ f^{-1}$ will not be conjugated to $\Gamma$
in ${\rm PSL}(2,\R)$, thus yielding Riemann surface $X^{\mu}=H/\Gamma^{\mu}$
as a deformation of $X$.

Specifically, infinitesimal deformations are described by sheaf
cohomology group $H^{1}(X,TX)$. Using Hodge theory (for the  choice of
Poincar\'{e} metric on $X$), one can identify it with the complex linear space
${\cal H}^{-1,1}(X)$ of harmonic Beltrami differentials on $X$. This introduces
a natural complex structure on Teichm\"{u}ller space ${\cal T}_{0,n}$,
with space ${\cal H}^{-1,1}(X)$ being its holomorphic tangent space at
point $X$. The inner product in ${\cal H}^{-1,1}(X)$, induced by the Hodge
$\ast$-operator is given by
\beq \label{IPWP}
(M,N)_{WP} \doteq \int_{X}M\bar{N} d\rho,
\eeq
and defines a Hermitian metric on ${\cal T}_{0,n}$. It is called the
Weil-Petersson metric, and it turns out to be K\"{a}hler~\cite{W,A2}.
The corresponding holomorphic cotangent space to ${\cal T}_{0,n}$ at point
$X$ is isomorphic to the linear space ${\cal H}^{2,0}(X)$ of harmonic
quadratic differentials on $X$. The pairing between these spaces---Serre
duality---is given by integration
\beq \label{P}
\int_{X}MQ,~M \in {\cal H}^{-1,1}(X)~Q \in {\cal H}^{2,0}(X),
\eeq
since the product of a Beltrami differential and a quadratic differential is a
$(1,1)$-form on $X$. According to (\ref{P}), Hermitian product (\ref{IPWP})in
the tangent space induces Hermitian product (\ref{IP}) in the space of
harmonic quadratic differentials.

These results generalize verbatim to the case of Riemann surfaces of the type
$(g,n)$---Riemann surfaces of genus $g$ with $n$ punctures, satisfying the
condition $\chi(X)=2-2g+n <0$. Corresponding Teichm\"{u}ller space
${\cal T}_{g,n}$ is a complex manifold of dimension $3g-3+n$ with a natural
K\"{a}hler structure, given by the Weil-Petersson metric.

In our case the presentation is simplified by considering the space of
punctures $Z_n$, introduced in \S2.2. It plays the role of the
``intermediate'' moduli space: it is covered by the Teichm\"{u}ller space
${\cal T}_{0,n}$ and in turn, covers the moduli space ${\cal M}_{0,n}$. The
complex linear space ${\cal H}^ {2,0}(X)$  can be described explicitly as
consisting of meromorphic functions on $\bar{X}={\bf P}^1$  with at
most simple poles at the punctures $z_1, \ldots, z_{n-1},$ having the order
$O(|z|^{-3})$ as $z \rightarrow z_n=\infty$. It is $n-3$-dimensional
and has a special basis $\{P_i\}$ associated with punctures,
$$P_{i}(z) \doteq -\frac{1}{\pi}R(z,z_i),~i=1, \ldots, n-3,$$
where $R(z,z_i)$ is given by (\ref{R}).
As we mentioned, in the cotangent space the Weil-Petersson metric has the form
\beq \label{WPQD}
(P,Q) = \int_{X} P(z) \overline{Q(z)}e^{-\phi_{cl}(z, \bar{z})}
d^{2}z,~P,Q \in {\cal H}^{2,0}(X).
\eeq
When lifted to upper half-plane $H$, it coincides with Petersson's inner
product for $\Gamma$-automorphic forms of weight $4$.

Next, denote by $\{ Q_i\}^{n-3}_{i=1}$ the basis in ${\cal H}^ {2,0}(X)$,
orthogonal to $\{ P_i\}^{n-3}_{i=1}$, i.e.
\beq \label{O}
(P_{i},Q_{j})=\delta_{ij}.
\eeq
Harmonic Beltrami differentials
\beq \label{HM}
M_{i}(z,\bar{z}) \doteq e^{-\phi_{cl}(z, \bar{z})} \overline{Q_{i}(z)}
\eeq
define quasi-conformal mappings $F^{\epsilon}_i$, which infinitesimally
``move'' only the puncture $z_{i}$ and keep other punctures fixed. Indeed,
as it follows from Beltrami equation, the infinitesimal deformation
$$\dot{F}_i=(\frac{\partial}{\partial \epsilon} F^{\epsilon}_i)|_{\epsilon=0},
$$
satisfies the $\bar{\pa}$-equation
$$\frac{\pa\dot{F}_i}{\pa \bar{z}}=M_i,$$
and, therefore, admits the following integral representation
\beq \label{INTR}
\dot{F}_i(z)=-\frac{1}{\pi}\int M(w)R(w,z)d^2w.
\eeq
Now from (\ref{INTR}) and (\ref{WPQD})---(\ref{HM}) it immediately follows that
\beq \label{FP}
\dot{F}_{i}(z_{j}) =\delta_{ij},
\eeq
which is a condition of infinitesimally moving only a given puncture
$z_{i}$. This shows that vector fields $\pa/\pa z_i$ on $Z_n$ at a point
corresponding to Riemann surface $X$, are represented by infinitesimal
deformations $\dot{F}_i,~i=1,\ldots,n-3$. As it also follows from (\ref{INTR}),
formula (\ref{FP}) can be specialized further as
\begin{eqnarray} \label{FP1}
\dot{F}_i(z)&=&\delta_{ij}+(z-z_j)\dot{F}_{iz}(z_j)+o(|z-z_j|),~z
\rightarrow z_j,~ j \neq n, \nonumber \\
\dot{F}_n(z)&=&z\dot{F}_{iz}(\infty)+o(|z|),~z \rightarrow z_n=\infty.
\end{eqnarray}

Finally, we describe the dependence of classical solution $\phi_{cl}(z;
z_{1}, \ldots ,z_{n-3})$ on the punctures. It is given by the following
``conservation law''
\beq \label{PHI}
\frac{\pa}{\pa z_i}(e^{\phi_{cl}})= \frac{\pa}{\pa z}(e^{\phi_{cl}}\dot{F}_i),
{}~i=1, \cdots, n-3,
\eeq
which is a reformulation of Ahlfors' result on vanishing of the first
variation of the Poincar\'{e} volume element under quasiconformal deformations
with harmonic Beltrami differentials~\cite{A2}.

\subsection{Proofs}

The reader can find the detailed account in~\cite[\S\S3-4]{ZT1}; here we just
highlight the main points.

In order to prove (\ref{AP}), one should use definition of Liouville action
(\ref{LA}), relation (\ref{TSW}), properties (\ref{FP1}) of vector
fields $\dot{F}_i$ and formula
$$\frac{\pa}{\pa z_i}\phi_{cl}=-(\phi_{cl})_z \dot{F}_i - \dot{F}_{iz},$$
which follows from conservation law (\ref{PHI}). Application of Stokes'
formula, with subsequent careful analysis of boundary terms as $\epsilon
\rightarrow 0$, results in relations (\ref{AP}).

In order to prove (\ref{APWP}), consider a one-parameter family of Riemann
surfaces $X^{\epsilon}_j=
\C \setminus\{z^{\epsilon j}_1, \cdots, z^{\epsilon j}_{n-3},0,1\},~j=1,
\ldots,n-3$, obtained by deformations $F^{\epsilon}_j$ for sufficiently
small $\epsilon$. The map $F^{\epsilon}_j$ satisfies Beltrami equation on $X$
$$\frac{\pa F^{\epsilon}_j}{\pa \bar{z}}=\epsilon M_j \frac{\pa F^{\epsilon}_j}
{\pa z},$$
and depends holomorphically on $\epsilon$; $z^{\epsilon j}_i=
F^{\epsilon}_j(z_i)$ are also holomorphic in $\epsilon$. Denote by
$J^{\epsilon}_j$ uniformization map for the Riemann surface $X^{\epsilon}_j$
and consider the following commutative diagram
$$
\begin{array}{rcl}
H & \stackrel{f^{\epsilon}_j}{\rightarrow} & H  \\
J  \downarrow & & \downarrow  J^{\epsilon}_j \\
X & \stackrel{F^{\epsilon}_j}{\rightarrow} & X^{\epsilon}_j
\end{array},
$$
where map $f^{\epsilon}_j$ is quasiconformal with Beltrami differential
$\mu=M_j \circ J \bar{J}^{\prime}/J^{\prime}$. Next, consider equation
$${\cal S}((J^{\epsilon}_j)^{-1} \circ F^{\epsilon}_j)=
{\cal S}(f^{\epsilon}_j \circ J^{-1}),$$
which follows from the commutative diagram, and use Caley identity (\ref{CI})
to obtain
\beq \label{EQ}
{\cal S}((J^{\epsilon}_j)^{-1}) \circ F^{\epsilon}_j~(F^{\epsilon}_{jz})^2
+{\cal S}(F^{\epsilon}_j)={\cal S}(f^{\epsilon}_j) \circ
J^{-1} (J^{-1^{\prime}})^2+{\cal S}(J^{-1}).
\eeq
Finally, apply partial derivative $\pa/\pa \bar{\epsilon}$ to equation
(\ref{EQ}) and evaluate the result at $\epsilon=0$
$$\sum_{i=1}^{n-3}\frac{\pa}{\pa \bar{\epsilon}} c^{\epsilon j}_i|_{\epsilon=0}
R(z,z_i)=\frac{\pa}{\pa \bar{\epsilon}}(f^{\epsilon}_{jzzz}
\circ J^{-1} (J^{-1^{\prime}})^2|_{\epsilon=0}.$$
Using relation
$$\frac{\pa}{\pa \bar{\epsilon}}f^{\epsilon}_{jzzz}|_{\epsilon=0}=
-\frac{1}{2}Q_j,$$
which is equivalent to the Ahlfors lemma in \S4.1, we get
$$\sum_{i=1}^{n-3}\frac{\pa c_i}{\pa \bar{z}_j}P_i=\frac{1}{2 \pi}Q_j,$$
so that
$$\frac{\pa c_i}{\pa \bar{z}_j}=\frac{1}{2 \pi}(Q_j,Q_i)=
\frac{1}{2 \pi}(\frac{\pa}{\pa z_i},\frac{\pa}{\pa z_j})_{WP}.$$

In order to prove CWI (\ref{EI}), apply partial derivative $\pa/\pa \epsilon$
to equation (\ref{EQ}) and evaluate the result at $\epsilon=0$
\beq \label{HDE}
\{\frac{\pa}{\pa z_i}+\dot{F}_i(z) \frac{\pa}{\pa z}+\dot{F}_{iz}(z)\}T_{cl}(z)
=\frac{1}{h}(\dot{f}_{i \zeta \zeta \zeta} \circ J^{-1}-\dot{F}_{izzz})(z).
\eeq
{}From the integral representation (\ref{INTR}) it follows that
\beq \label{MF}
\dot{f}_{i \zeta \zeta \zeta}(z)-\dot{F}_{izzz}(z)=2 \int_{X}\{
\frac{3}{\pi(z-w)^4}-{\cal D}_z {\cal D}_w G(z,w)\}e^{-\phi_{cl}(w)}
\overline{Q_i(w)}d^2w.
\eeq
Now (\ref{EI}) follows from formulas (\ref{L}), (\ref{HDE})---(\ref{MF}) and
orthogonality condition (\ref{O}).

We will not present here the proofs of the one-loop results~\cite{T1}. The
reader can find necessary technical tools in \cite{ZTIT}.

\subsection{Generalizations}

First, CWI for multi-point correlation functions also carry meaningful
information about Weil-Petersson geometry. Thus, for instance, CWI for
four-point correlation function $<<T(z_1)T(z_2)\bar{T}(\bar{w}_1)\bar{T}(
\bar{w}_2)X>>$ at the tree level reproduces Wolpert's result on evaluation of
Riemann tensor of the Weil-Petersson metric~\cite{Wol}. Moreover,
universal CWI (\ref{UW1})---(\ref{UW2}) carry, in a ``compressed form'',
important information about the modular geometry; the challenging problem is to
``decode'' it without appealing to perturbation theory.

Second, our approach can be trivially generalized to the case of Riemann
surfaces with the branch points of orders $l_i$, $2 \leq l_i \leq \infty$.
Corresponding changes in the definition of Liouville action (\ref{LA}) are
obvious. One has for the Schwarzian derivative of uniformization map (see,
e.g.,
\cite{P1})
$$T_{cl}(z)=\frac{1}{h}{\cal S}(J^{-1})(z)=\sum_{i=1}^{n}\{\frac{\alpha_i}{
(z-z_i)^2}+ \frac{c_i}{z-z_i}\},$$
where $\alpha_i=(1-l_{i}^{-2})/2h$ are classical dimensions $\Delta_i$
of geometric vertex operators $V_{\alpha_i}$ (recall (\ref{VC})). The
arguments,
used in the punctured case, apply to this case as well and show that after the
quantization conformal dimensions $\Delta_i$ remain the same.

Third, our approach can be generalized to Riemann surfaces of type $(g,n)$. In
this case (see~\cite{ZT2} for details when $n=0$), one can use single global
coordinate on the Schottky cover of Riemann surface $X$, provided by
Schottky uniformization. In its terms it is possible to define Liouville
action, expectation value $<X>$ by functional integral (\ref{FI}) and
generating functional (\ref{Z}). On Riemann surface $X$ this amounts to the
choice of Schottky projective connection $T^{S}$ and considering the
difference $T(\phi)-T^{S}/h$, which is well-defined quadratic differential
on $X$ (cf.~\cite{Son}).

Finally, we make the following speculative remark. In derivation~\cite{Pol1}
of Liouville action from $D$-dimensional bosonic string, the following
correspondence between Liouville's coupling constant $h$ and dimension $D$
was established
$$\frac{1}{2 \pi h}=\frac{25-D}{24\pi}$$
(note the replacement $26 \mapsto 25$, which should be considered as a
quantum correction). According to our calculations,
$$c_{Liouv}=\frac{12}{h}+1=25-D+1=26-D.$$
This expression, when added to contribution $D$ from the string modes and
to the contribution $-26$ coming from Faddeev-Popov ghosts, yields zero and
thus cancels the global conformal anomaly for any $D$.

{\bf Acknowledgments} I would like to thank the organizers and the staff of
the International School of Physics ``Enrico Fermi'', held at Varenna, Villa
Monastero, 28 June -- 8 July 1994, for their kind hospitality. This work was
partially supported by the NSF grant DMS-92-04092.

\end{document}